\pdfoutput=1 %for arXiv submission

\documentclass[twocolumn]{aastex62}
\usepackage{hyperref}
\usepackage{amsmath,amstext}
\usepackage[T1]{fontenc}
\usepackage{natbib}
\usepackage[figure,figure*]{hypcap}
\usepackage{graphicx}
\usepackage{tablefootnote}
\usepackage{units}
\usepackage{multirow}
\usepackage{tabularx}

\def\sun{$_{\odot}$}

\def\f28{${f}_{2-8{\rm keV}}$}

\def\sun{$_{\odot}$}

\def\ergscm2{erg s$^{-1}$ cm$^{-2}$}
\def\yr-1{yr$^{-1}$}

\def\asec{\ifmmode^{\prime\prime}\else$^{\prime\prime}$\fi}

\received{date}
\submitjournal{ApJ}

\shorttitle{CL Quasar Hosts} 
\shortauthors{Charlton et al.}

\begin{document}

\title{Gemini Imaging of the Host Galaxies of Changing-Look Quasars}

\correspondingauthor{Paul J. L. Charlton}
\email{paul.charlton@mail.mcgill.ca}

\author{Paul J. L. Charlton}
\affil{McGill Space Institute and Department of Physics, McGill University, 3600 rue University, Montreal, Quebec, H3A 2T8, Canada}
\author{John J. Ruan}
\affil{McGill Space Institute and Department of Physics, McGill University, 3600 rue University, Montreal, Quebec, H3A 2T8, Canada}
\author{Daryl Haggard}
\affil{McGill Space Institute and Department of Physics, McGill University, 3600 rue University, Montreal, Quebec, H3A 2T8, Canada}
\affil{CIFAR Azrieli Global Scholar, Gravity \& the Extreme Universe Program, Canadian Institute for Advanced Research, 661 University Avenue, Suite 505, Toronto, ON M5G 1M1, Canada}
\author{Scott F. Anderson}
\affil{Department of Astronomy, University of Washington, Box 351580, Seattle, WA 98195, USA}
\author{Michael Eracleous}
\affil{Department of Astronomy and Astrophysics and Institute for Gravitation and the Cosmos, The Pennsylvania State University, 525 Davey Lab, University Park, PA 16803, USA}
\author{Chelsea L. MacLeod}
\affil{Harvard Smithsonian Center for Astrophysics, 60 Garden St, Cambridge, MA 02138, USA}
\author{Jessie C. Runnoe}
\affil{Department of Astronomy, University of Michigan, 1085 S. University Avenue, Ann Arbor, MI 48109, USA}

\begin{abstract}
Changing-look quasars are a newly-discovered class of luminous active galactic nuclei that undergo rapid ($\lesssim$10 year) transitions between Type 1 and Type 1.9/2, with an associated change in their continuum emission. We characterize the host galaxies of four faded changing-look quasars using broadband optical imaging. We use \textit{gri} images obtained with the Gemini Multi Object Spectrograph (GMOS) on Gemini North to characterize the surface brightness profiles of the quasar hosts and search for [O III] $\lambda4959,\lambda5007$ emission from spatially extended regions, or voorwerpjes, with the goal of using them to examine past luminosity history. Although we do not detect, voorwerpjes surrounding the four quasar host galaxies, we take advantage of the dim nuclear emission to characterize the colors and morphologies of the host galaxies. Three of the four galaxies show morphological evidence of merger activity or tidal features in their residuals. The three galaxies which are not highly distorted are fit with a single S\'ersic profile to characterize their overall surface brightness profiles. The single-S\'ersic fits give intermediate S\'ersic indices between the $n=1$ of disk galaxies and the $n=4$ of ellipticals. On a color-magnitude diagram, our changing-look quasar host galaxies reside in the blue cloud, with other AGN host galaxies and star-forming galaxies. On a color-S\'ersic index diagram the changing-look quasar hosts reside with other AGN hosts in the "green valley". Our analysis suggests that the hosts of changing-look quasars are predominantly disrupted or merging galaxies that resemble AGN hosts, rather than inactive galaxies.
\end{abstract}

% ============================
\section{Introduction}
\label{sec:intro}

In the current galaxy evolution paradigm, quasars are active galactic nuclei (AGN) associated with rapid growth of the supermassive black hole (SMBH), during brief periods ($10^{7-8}$ year) in a galaxy's life \citep{2001ApJ...547...12M, 2004MNRAS.351..169M, 2002MNRAS.335..965Y}. This rapid mass growth occurs through a luminous accretion disk, which emits a characteristic spectrum that peaks in the far UV in unobscured quasars \citep{1964ApJ...140..796S, 1969Natur.223..690L, 1984ARA&A..22..471R}.

The primary optical characteristics of quasars are their blue continua, and various emission lines. Type-1 quasars are high-luminosity AGN which show broad Balmer lines in their rest-frame optical spectra. These lines are associated with virialized, high-velocity gas in the broad line region (BLR). The spectra of these objects also contain narrow emission lines due to gas located in the narrow line region (NLR) at larger distances from the central engine. Type-2 quasars are similar, but are missing broad lines, and are less optically luminous than their Type-1 counterparts. Intermediate classes also exist; such as Type-1.8 and 1.9 which show weak or no broad H$\beta$ lines, while retaining broad H$\alpha$ emission.  

Efforts to unify Type-1 and 2 quasars have often led to an orientation-based explanation. In this scheme, a dusty torus or region of clumpy dust clouds of sufficient column density obscures the BLR along our line of sight; the NLR, being further away remains unobscured \citep{1985ApJ...297..621A, 1993ARA&A..31..473A, 1995PASP..107..803U}. Intermediate classes are explained via partially obscured BLRs or reddening (e.g., \citealt{1981ApJ...249..462O, 1995ApJ...454...95M, 2012MNRAS.426.2703S}). 

This simple orientation-based scheme has been challenged by evidence indicating that accretion rate plays a role in determining a quasar's type (e.g. \citealt{2004MNRAS.352.1390M}). Long and short-term studies of quasars have revealed significant variability in their light curves on all timescales and at all wavelengths likely driven by accretion rate variations (e.g., \citealt{1997ARA&A..35..445U, 2010ApJ...721.1014M, 2014ApJ...788...48S}). The recent discovery of changing look (CL) quasars offers evidence in support of a more complex picture of AGN unification put forward by \cite{1984MNRAS.211P..33P}. CL quasars exhibit extreme variability and appear to transition from Type-1 to Type-2 (or Types-1.8/1.9), or vice versa, a behavior not observed in standard quasars. This change in the broad lines is coupled with a significant decrease (or increase) in continuum emission, on timescales of $\lesssim10$ years \citep{2015ApJ...800..144L, 2016MNRAS.455.1691R, 2016ApJ...826..188R, 2016MNRAS.457..389M, 2016MNRAS.461.1927P, 2016A&A...593L...8M, 2017ApJ...835..144G,  2017ApJ...849..102C, 2018ApJ...858...49W, 2018ApJ...862..109Y, 2018ApJ...864...27S, 2018arXiv180506921R}. 

%Before considering a rapid fading of the quasar, explanations of this phenomenon via extrinsic factors must be considered. These other explanations include invoking large dust clouds obscuring or revealing the BLR over repeated observations, and tidal disruption events (TDE) in which a star is ripped apart by tidal forces, generating a bright flare. Episodic obscuration of the BLR is at odds with the long dynamical timescales required for a cloud in a Keplerian orbit outside of the BLR to block it \citep{2015ApJ...800..144L}. Modeling the changes in the spectrum via extinction also fails to explain the broadening of the broad emission lines in the faint state \citep{2016ApJ...826..188R} and the observations showing that the IR luminosity decreases along with the optical luminosity \citep{2018ApJ...864...27S, 2018ApJ...862..109Y}. \cite{2015MNRAS.452...69M} suggest TDEs can be responsible for some CL quasars' behavior by briefly re-exciting the gas around a faded quasar. This explanation is disfavored by the inability of a single TDE to sufficiently ionize the NLR to the degree that is observed, populate the BLR with enough gas to generate the observed emission, as well as clear evidence for persistent AGN in all CL quasars.

Before considering a rapid fading of the quasar, explanations of this phenomenon via extrinsic factors must be considered. These other explanations include invoking large dust clouds obscuring or revealing the BLR over repeated observations, and tidal disruption events (TDE) in which a star near the vicinity of black hole is torn apart by tidal forces, leading to a bright flare. Episodic obscuration of the BLR is at odds with the long dynamical timescales required for a cloud in a Keplerian orbit outside of the BLR to block it \citep{2015ApJ...800..144L}. Modeling the changes in the spectrum via extinction also fails to explain the broadening of the broad emission lines in the faint state \citep{2016ApJ...826..188R} and the observations showing that the IR luminosity decreases along with the optical luminosity \citep{2018ApJ...864...27S, 2018ApJ...862..109Y}. \cite{2015MNRAS.452...69M} suggest TDEs can be responsible for some CL quasars' behavior by briefly re-exciting the gas around a faded quasar. There is clear evidence for persistent AGN in all CL quasars, but having the ability to examine variability in the distant past would allow us to distinguish TDE from intrinsic fading. If we could determine that the quasar was bright on timescales of $~10^5$ years, the observed fading is less likely to be due to a TDE.

%The lack of satisfying alternative explanations for CL quasars' transitions forces us to consider that there is an intrinsic change in the BLR emission between the bright state and faint state over short timescales. There is observed to be considerable broadening of the broad Balmer emission as the quasar fades. The BLR emission is due to recombination of ionized Hydrogen. As the quasar luminosity decreases, the ionizing flux decreases accordingly. Consequently, the regions able to be ionized lie closer to the SMBH than when the quasar is in its brighter state. Due to the faster orbital velocities of the closer gas, the broad line emission broadens relative to when the quasar is in the bright state. 

One method of determining whether these systems were indeed long-lived quasars in their bright state (rather than a short timescale phenomenon) is to look at their local environment for signs of past quasar activity. UV photons from the inner regions of a quasar accretion disk can photoionize the surrounding gas \citep{1981PASP...93....5B, 1987ApJS...63..295V, 2003MNRAS.346.1055K, 2006MNRAS.372..961K}. The UV emission of particularly luminous AGN can ionize gas out to kpc scales, creating large NLR (e.g. in [O III] $\lambda5007$) known as extended emission line regions \citep[EELR; e.g.][and more recently, \citealt{2018MNRAS.480.2302S}]{1985ApJ...293..120B, 1987ApJ...316..584S}. The size of the EELR (and hence strength of [O III] emission) have been shown to correlate with the luminosity of the AGN \citep{2002ApJ...574L.105B,2013ApJ...774..145H,2014ApJ...787...65H}. %These EELRs can be connected to the quasar host galaxy or in the surrounding environment. 

In the context of CL quasars, one promising avenue of investigation is searching for voorwerpjes or quasar ionization echoes, named for the first to be discovered, Hanny's Voorwerp \citep{2009MNRAS.399..129L, 2010ApJ...724L..30S, 2010A&A...517L...8R, 2012AJ....144...66K}. Voorwerpjes are EELRs glowing brightly in [O III] and can span tens of kpc from the galaxies, that contain no apparent quasar. Recent work indicates that voorwerpjes are typically found around faded quasars \citep{2015AJ....149..155K, 2017ApJ...835..256K}, and can be used to trace the luminosity and variability of the central engine over $10^{4-5}$ years. We can take advantage of the fact that CL quasars have faded from Type 1 to Type 2 to explore their host galaxies and surrounding environment. Furthermore, finding voorwerpjes around faded CL quasars may allow us to definitively rule out other explanations of the changing-look phenomenon. 

If CL quasars are a distinct subclass of AGN, or represent a particular state in a quasar's life, characterizing the host galaxies could play an important role in understanding galaxy evolution. Previous photometric studies of Type 1 quasar host galaxies have been forced to model and subtract the bright AGN and/or restrict themselves to low redshifts (e.g., \citealt{1999MNRAS.308..377M, 2001ApJ...555..719C, 2004ApJ...614..586S, 2014ApJ...780..162M}). In contrast, the faded central engines of CL quasars afford us a clear view of the host galaxy itself, allowing us to not only search for voorwerpjes, but accurately characterize the color, luminosity, and morphology of the host galaxies with minimal contamination from the AGN.

In this work we investigate a sample of four changing look quasars using high resolution broadband optical imaging with Gemini North's Gemini Multi-Object Spectrograph (GMOS). This approach allows us to characterize the morphological and photometric properties of the quasar host galaxies, and, via subtraction of the stellar continuum, look for extended emission line regions or other morphological features. In Section \ref{sec:data} we describe our target quasars and summarize our observations. Section \ref{sec:meth} covers our data reduction procedure and our process for subtracting the host galaxy continuum emission. Section \ref{sec:results} contains our galaxy models and continuum-subtracted images, along with discussion of residual features for each galaxy. Section \ref{sec:disc} contains discussion of features observed in the CL quasar host galaxies and the conclusions that can be drawn from our analysis. 

Throughout this paper we assume a flat $\Lambda\mathrm{CDM}$ cosmology: $H_0 = 70$ $\mathrm{km}\ \mathrm{s}^{-1}\mathrm{Mpc}^{-1}$, $\Omega_{m} = 0.3$, $\Omega_{\Lambda} = 0.7$.

% ====================================
\section{Data}
\label{sec:data}
\subsection{Target Selection}
\label{subsec:targets}
In this paper, we investigate four turn-off CL quasars, which have lost their broad emission lines and faded considerably in repeat SDSS \citep{2000AJ....120.1579Y} spectra. The quasars are SDSS J015957.64+003310.5 \citep[hereafter J0159+0033;][]{2015ApJ...800..144L}, SDSS J101152.90+544206.4 \citep[hereafter J1011+5442;][]{2016MNRAS.455.1691R}, SDSS J012648.08$-$083948.0, and SDSS J233602.98+001728.7 \citep[hereafter J0126$-$0839 and J2236+0017;][]{2016ApJ...826..188R}. Photometric and spectroscopic information, along with stellar masses for he host galaxies of these quasars are listed in Table \ref{tab:lit_summ}\\

\begin{deluxetable*}{ccccccccccc} 
\tabletypesize{\footnotesize} 
\tablecolumns{11} 
\tablewidth{0pt} 
\tablecaption{Observed CL Quasar Properties} 
\tablehead{\colhead{Quasar Name} & \colhead{$z_{\mathrm{spec}}$} & \colhead{$M_{*}$} & \colhead{$\mathrm{MJD}_{\mathrm{Phot}}$} & \colhead{$m_g$} & \colhead{$m_r$} & \colhead{$m_i$} & \colhead{$\mathrm{MJD}_{\mathrm{Spec}}$} & \colhead{$\mathrm{log}\left(\lambda L_{5100}\right)$} & \colhead{Type} & \colhead{Ref} \vspace{-0.2cm}\\
    \ & & \colhead{$(10^{10} M_{\odot})$} & & & & & & \colhead{$(\mathrm{erg} \mathrm{s}^{-1})$} & &}
\startdata
    \ J0126$-$0839 & 0.198 & $3.5_{2.5}^{4.9}$ & 51813 & $18.45$ & $18.03$ & $17.63$ & 52163 & $43.43\pm0.03$ & 1 & \cite{2016ApJ...826..188R}\\
    \  & & & & $\pm0.01$ & $\pm0.01$ & $\pm0.01$ & & & & \\
    \ & & & 56016 & $19.55$ & $18.81$ & $18.43$ & 54465 & $<42.30$ & 2 & \\
    \ & & & & $\pm0.02$ & $\pm0.02$ & $\pm0.04$ & & & & \\
    \ J0159+0033 & 0.312 & $4.7_{3.9}^{8.0}$ & 52963 & $19.42$ & $18.85$ & $18.52$ & 51871 & $43.52\pm0.05$ & 1 & \cite{2015ApJ...800..144L}\\
    \  & & & & $\pm0.01$ & $\pm0.01$ & $\pm0.01$ & & & & \\
    \ & & & 56316 & $19.96$ & $19.21$ & $18.99$ & 55201 & $43.27\pm0.06$ & 1.9 & \\
    \ & & & & $\pm0.07$ & $\pm0.02$ & $\pm0.02$ & & & & \\
    \ J1011+5442 & 0.246 & $1.01_{0.93}^{1.09}$ & 52318 & $18.12$ & $17.88$ & $17.63$ & 52652 & $43.80\pm0.01$ & 1 & \cite{2016MNRAS.455.1691R}\\
    \  & & & & $\pm0.01$ & $\pm0.01$ & $\pm0.01$ & & & & \\
    \ & & & 56265 & $19.771$ & $19.19$ & $18.84$ & 57073 & $42.80\pm0.01$ & 2 & \\
    \ & & & & $\pm0.02$ & $\pm0.04$ & $\pm0.03$ & & & & \\
    \ J2336+0017 & 0.243 & $1.8_{1.5}^{2.5}$ & 53288 & $19.64$ & $18.99$ & $18.60$ & 52096 & $43.04\pm0.09$ & 1 & \cite{2016ApJ...826..188R}\\
    \  & & & & $\pm0.02$ & $\pm0.02$ & $\pm0.01$ & & & & \\
    \ & & & 56132 & $20.55$ & $19.88$ & $19.55$ & 55449 & $42.56\pm0.18$ & 1.9 & \\
    \ & & & & $\pm0.05$ & $\pm0.02$ & $\pm0.02$ & & & & \\
\enddata 
\tablecomments{Summary of observed quasar properties. Stellar masses were determined using a Salpeter IMF and STELIB library in the Value Added Catalog of \cite{2017arXiv171106575C}. The magnitudes in the bright state are taken from SDSS DR9 and converted to Pan-STARRS1 magnitudes using the conversion described in \cite{2015ApJ...806..244M} for consistency with observations in the faint state. Each quasar has two rows, one in the bright state, followed by one in the faint state.}
\label{tab:lit_summ}
\end{deluxetable*}

\subsection{Observations}
\label{subsec:obs}
Our data consist of observations taken with GMOS in imaging mode at Gemini North between September and October of 2017. Each of the four CL quasars was imaged in the broadband \textit{g}, \textit{r}, and \textit{i} filters. SDSS magnitudes of nearby stars were used for photometric calibration. The modeled magnitude of each component of the quasar host galaxies are determined based on this calibration.

The science data were reduced using the Gemini package for \textit{PyRAF}\footnote{https://www.gemini.edu/node/11823}. The GMOS image data reduction process consisted of using \texttt{gbias} to create the bias frame and \texttt{giflat}\ to create the normalized flat-field frame using twilight flats. We then use \texttt{gireduce} to subtract the overscan level, trim the science image, subtract the bias frame, and divide by the flat-field, and \texttt{imcoadd} to combine sub-exposures and science images. The final images are composed of a minimum of six exposures of 120 seconds each, summarized in Table \ref{tab:obs_summ}. At least five exposures per band, per CL quasar, were performed with seeing $~0.5-0.6\asec$. Rather than exclude exposures with relatively poor seeing, we opt to include as many exposures as possible to search for faint features, such as tidal streams, diffuse stellar halos, and faint voorwerpjes and EELRs.

Three-color GMOS images of the CL quasar host galaxies are shown in Figure \ref{fig:3_colour}, along with the SDSS spectra as observed in their faint states. The quasars selected for this work contain narrow [O III] emission lines in the \textit{r}-band, which allow us to compare between our three bands to search for EELRs and voorwerpjes. Figure \ref{fig:3_colour} also contains the residual images created by subtracting off the continuum emission; these residuals will be discussed in detail in Section \ref{sec:results}.

\begin{deluxetable}{cccccc} 
\tabletypesize{\footnotesize} 
\tablecolumns{6} 
\tablewidth{0pt} 
\tablecaption{Summary of Gemini North Observations} 
\tablehead{\colhead{Quasar Name} & \colhead{Band} & \colhead{Coadds} & \colhead{$\left\langle \mathrm{Seeing}\right\rangle$} & \colhead{Zero} & \colhead{Scale} \vspace{-0.2cm}\\ 
 &  &  & \colhead{$\left(\asec\right)$} & \colhead{Point} & \colhead{$\left(\nicefrac{\mathrm{kpc}}{\asec}\right)$} }
\startdata
J0126$-$0839 & \textit{g} & 7 & $0.65\pm0.02$ & 28.05 & 3.247\\
 & \textit{r} & 7 & $0.54\pm0.09$ & 28.36 & \\
 & \textit{i} & 7 & $0.53\pm0.09$ & 28.29 & \\
J0159+0033 & \textit{g} & 7 & $0.54\pm0.01$ & 28.18 & 4.576\\
 & \textit{r} & 6 & $0.54\pm0.08$ & 28.42 &\\
 & \textit{i} & 11 & $0.6\pm0.1$ & 28.46 &\\
J1011+5442 & \textit{g} & 6 & $0.6\pm0.1$ & 28.92 & 3.864\\
 & \textit{r} & 6 & $0.5\pm0.1$ & 28.62 & \\
 & \textit{i} & 6 & $0.49\pm0.09$ & 28.60 & \\
J2336+0017 & \textit{g} & 6 & $0.53\pm0.02$ & 28.18 & 3.829\\
 & \textit{r} & 13 & $0.5\pm0.1$ & 28.62 & \\
 & \textit{i} & 14 & $0.6\pm0.2$ & 28.60 & \\
\enddata 
\tablecomments{The average seeing, magnitude zero-point, and physical scale of each of our quasar host galaxy images. The average seeing of each image was determined by running the \textit{gemseeing} task on the final coadd, and the magnitude zeropoints were determined using SDSS photometric standard stars.}
\label{tab:obs_summ}
\end{deluxetable}

% ====================================
\section{Image Analysis}
\label{sec:meth}
We wish to isolate any possible extended photoionized [O III] emission-line regions. To accomplish this, we model and subtract the stellar continuum and nuclear emission in each of our broadband science coadds using \textit{GALFIT3} \citep{2002AJ....124..266P, 2010AJ....139.2097P} and \textit{GALFITM} \citep{MegaMorph}. 

\subsection{S\'ersic Profile}
The S\'ersic profile \citep{1968adga.book.....S} is an empirically-determined surface brightness profile that fits both disk and elliptical galaxies through the use of a variable index $n$ known as the S\'ersic index. The projected S\'ersic profile as a function of radius $r$ is given by
\begin{equation}
	\Sigma(r) = \Sigma_{\mathrm{eff}}\exp\left\{ \kappa\left[\left(\frac{r}{r_\mathrm{eff}} \right)^\frac{1}{n} -1 \right] \right\}
\end{equation}
where $r_{\mathrm{eff}}$ is the half-light radius, $\Sigma_{\mathrm{eff}}$ is the surface brightness at the half light radius and $\kappa$ is a parameter determined by $n$ and is not free.

The S\'ersic index is a useful parameter for the purpose of classifying the morphologies of our galaxies. Typically, bulge dominated ellipticals have high S\'ersic indices $\geqslant4$ which creates a surface brightness profile that is very centrally concentrated and tapers-off gradually. Disk-dominated spiral galaxies have S\'ersic indices closer to $\sim1$, which creates a profile that is relatively flat inside of the effective radius and drops off dramatically outside to create a relatively hard "edge" to the light profile.

\subsection{Galaxy Fitting}

The fits produced by \textit{GALFIT} can depend strongly on the PSF and sky background level if care is not taken to model them correctly. Although these quasars have faded, they have (usually) not disappeared completely. This residual quasar continuum emission, along with some host galaxy light, must be modeled with a PSF to avoid biasing the S\'ersic profiles toward higher central concentrations. For our PSFs, we used nearby stars on the same detector. Through our tests, we determined that using an infinite signal-to-noise model PSF produced by stacking multiple Gaussians did not significantly improve our galaxy fits, so instead we elected to use the sky-subtracted star cutouts as the PSF to model the point-like quasar emission in our final multi-component and overall fits to our quasar host galaxies. In order to model the sky background we include a significant number of background pixels. Of the total cutout size of $300\times300$ pixels, less than 3\% contain the CL quasar host galaxy, but we mask it out to avoid its flux raising the overall level of the sky background. Other galaxies in the cutout region are masked out using the segmentation map generated by \textit{SourceExtractor}\footnote{https://www.astromatic.net/software/sextractor}. We fit the sky with a central value and a gradient in the x and y directions. From this point, we can move on to fitting the host galaxies themselves.

As small-scale structures are of interest to us, we limit our galaxy fits to large-scale features, i.e., disks, bulges, extended stellar halos, and the quasar itself, so as not to remove features of interest from the residuals. For this purpose we use two S\'ersic components per galaxy (with the notable exception of J1011+5442). %We also include a PSF component in our \textit{GALFIT} model to account for the nuclear point source. In each quasar there remains some point-like emission that can interfere with a S\'ersic only fit by artificially increasing the S\'ersic index to increase the relative emission from the core of the quasar host galaxy.

After performing multi-component fits to remove as much of the stellar continuum as possible, we perform a single S\'ersic fit holding the sky at the level determined in the multi-S\'ersic fit. This allows us to determine an overall half light radius, and assess whether the galaxy appears to be bulge or disk dominated using the S\'ersic index.

\section{Results}
\label{sec:results}

\begin{figure*}
	%\centering
	\includegraphics[width=0.6\textwidth]{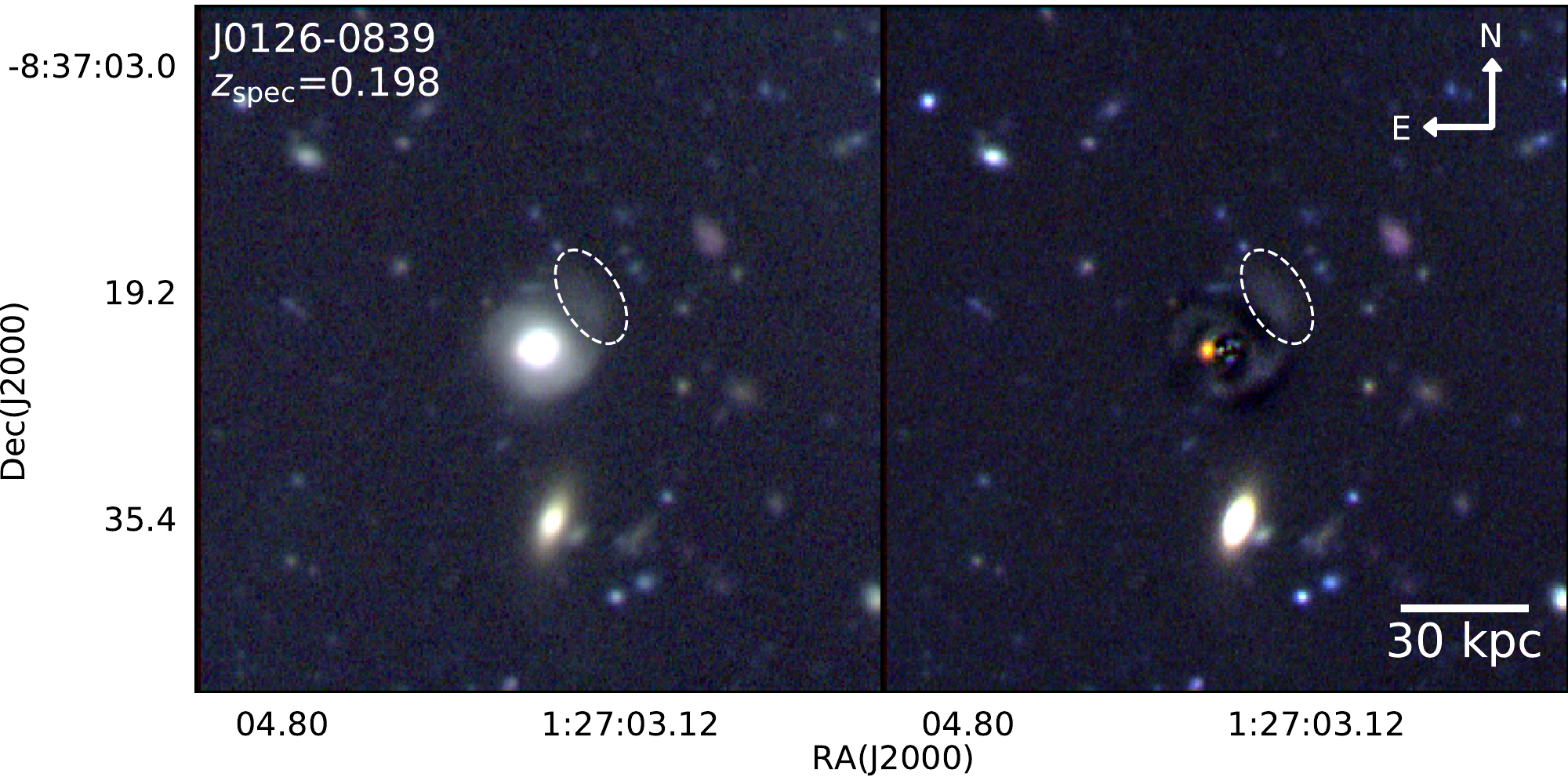}\includegraphics[width=0.4\textwidth]{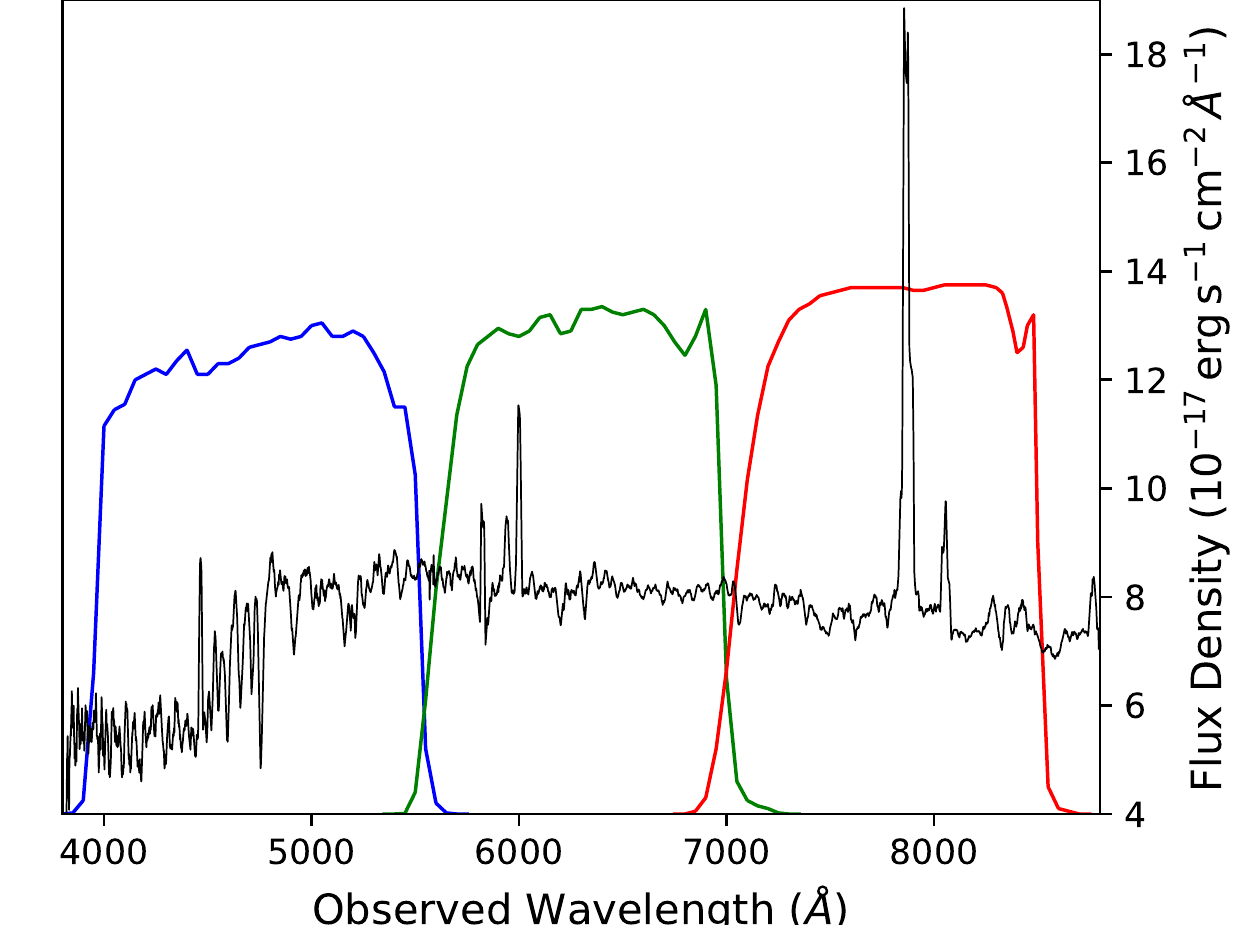}
    \includegraphics[width=0.6\textwidth]{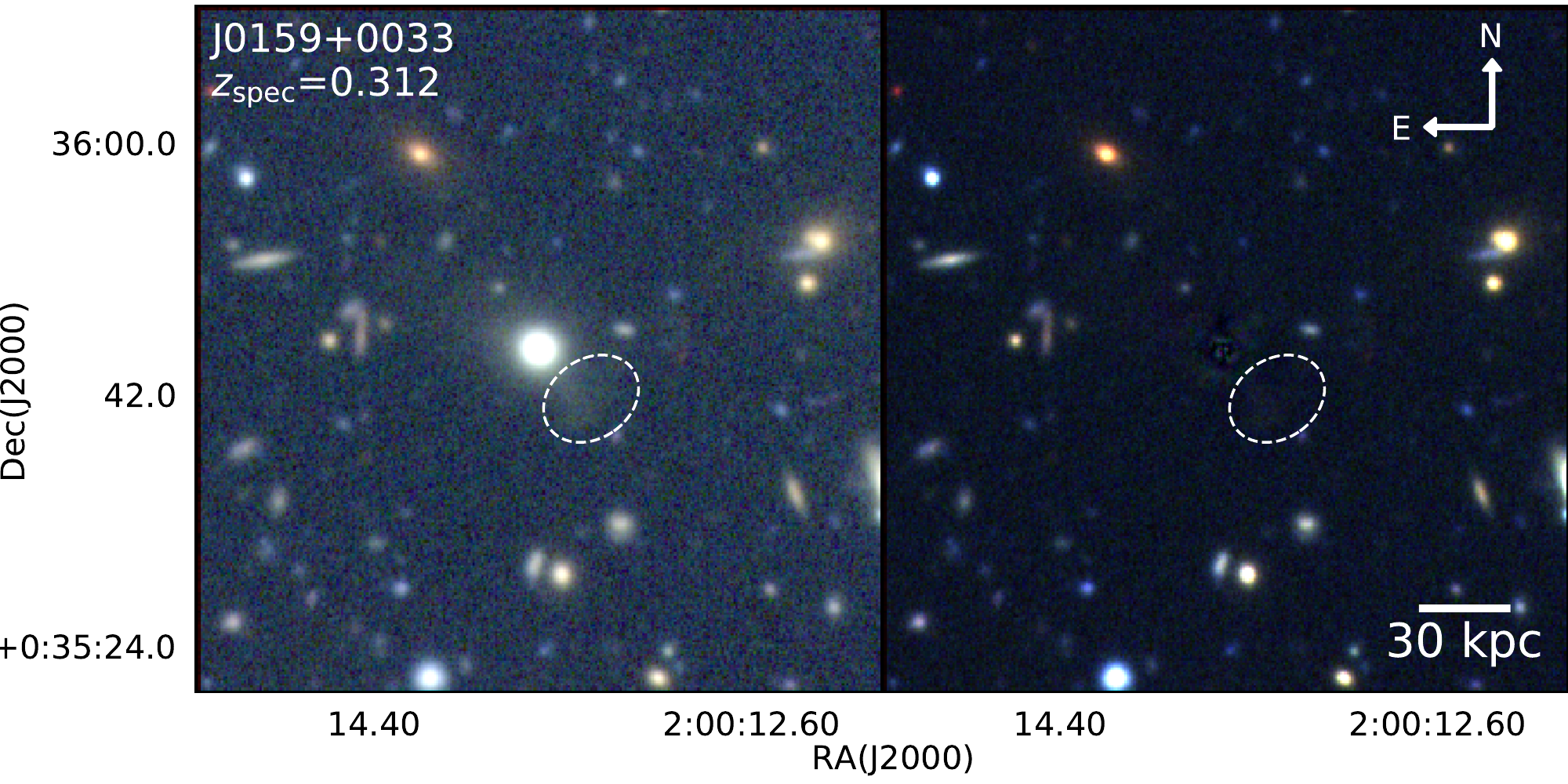}\includegraphics[width=0.4\textwidth]{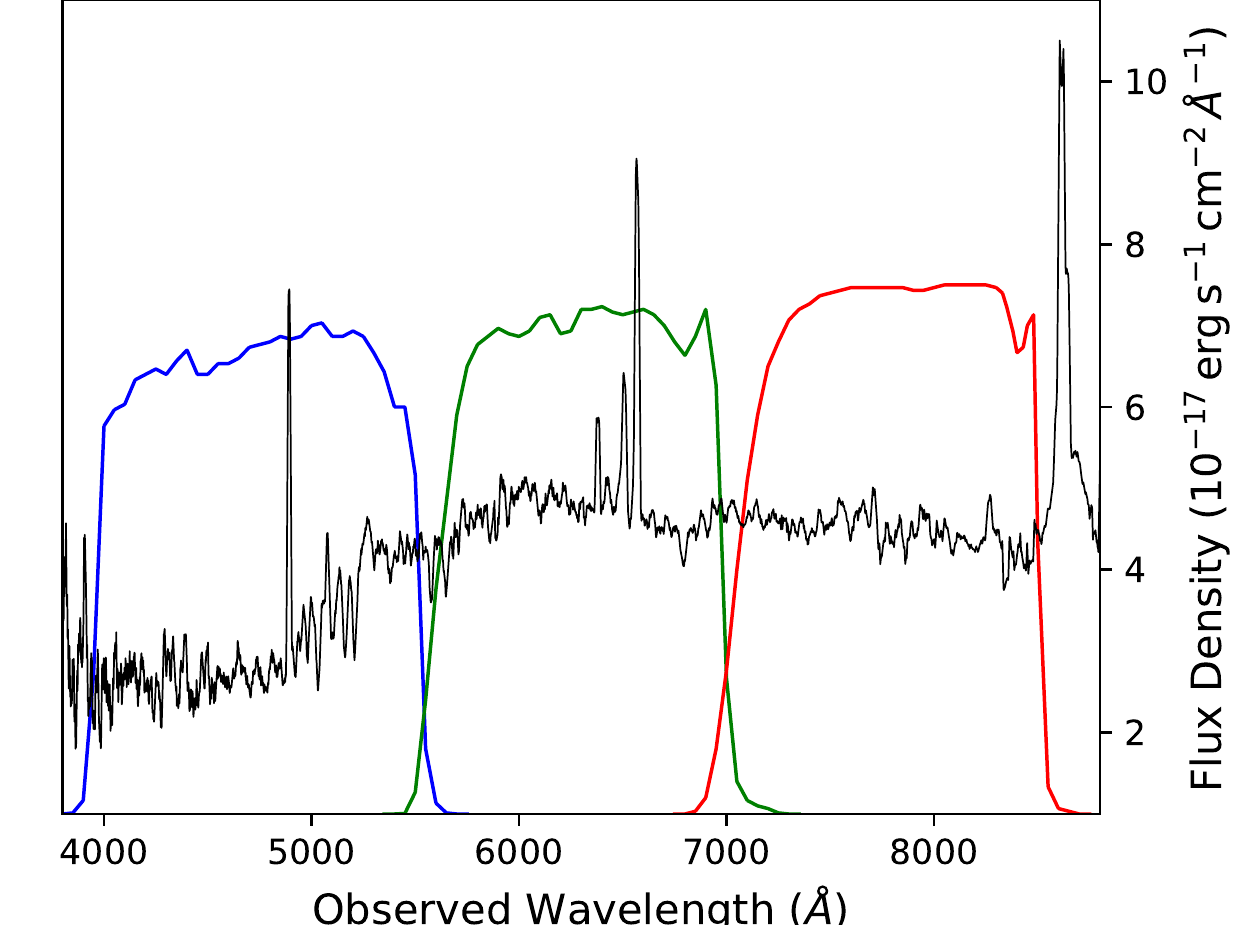}
    \includegraphics[width=0.6\textwidth]{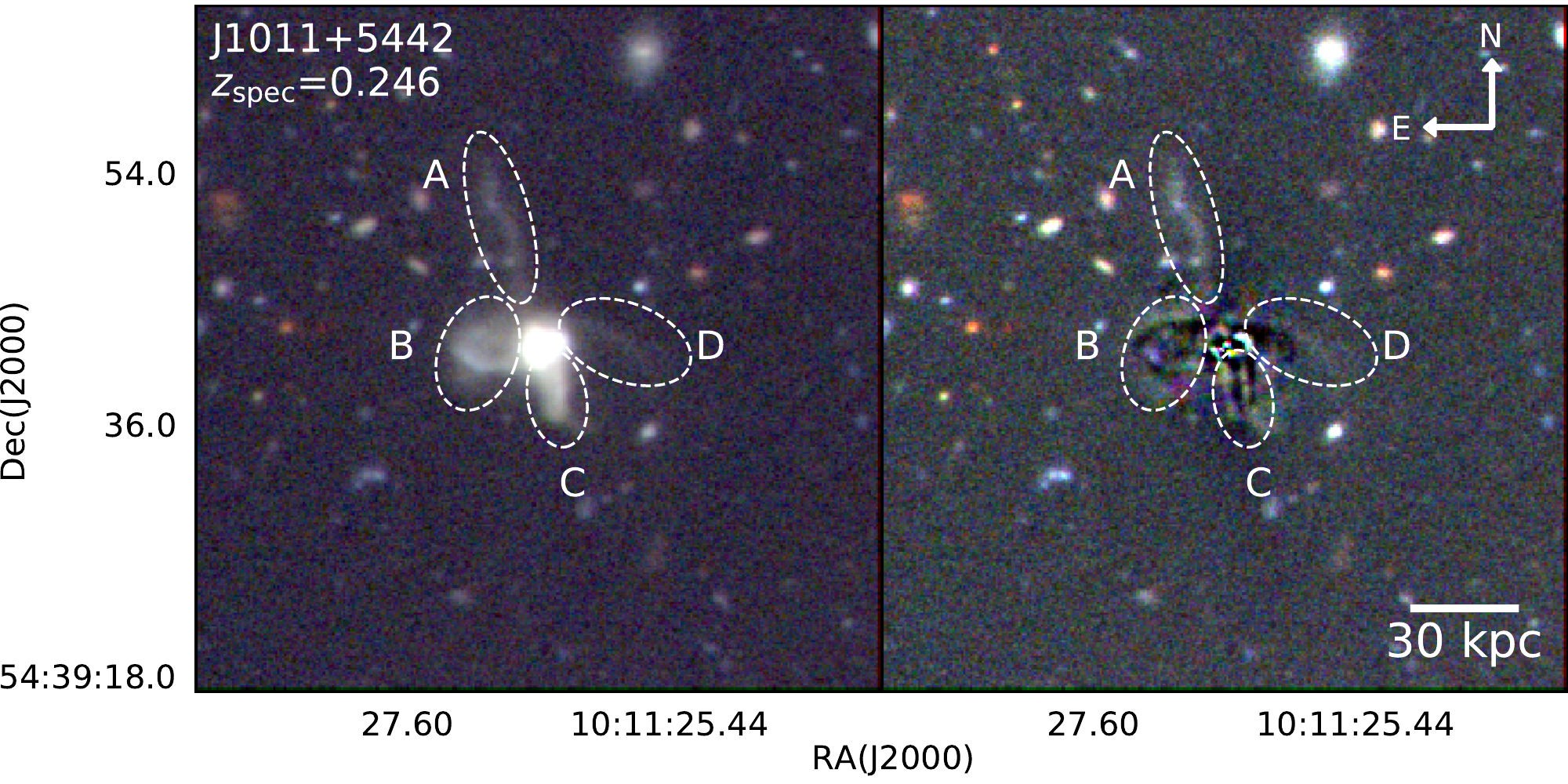}\includegraphics[width=0.4\textwidth]{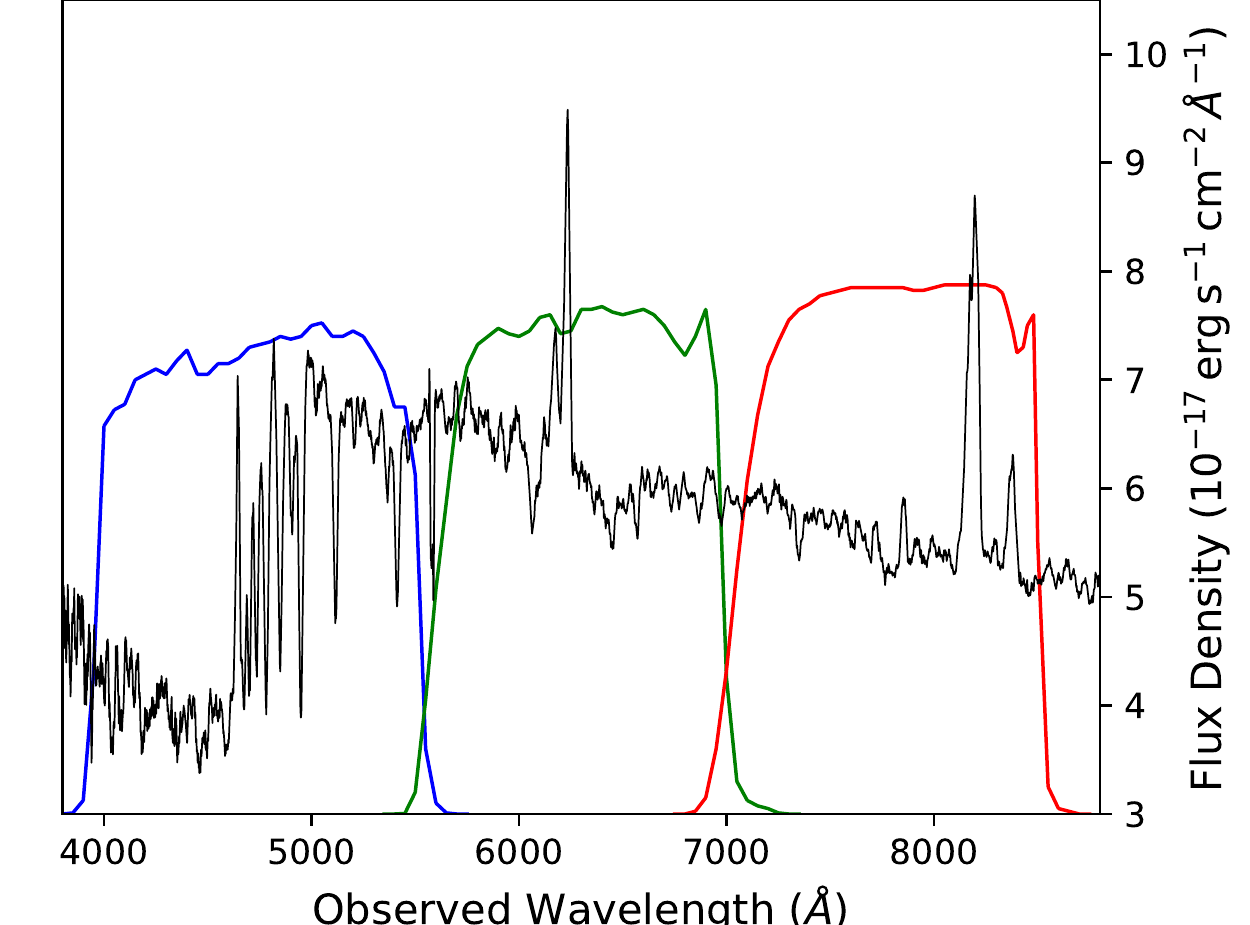}
    \includegraphics[width=0.6\textwidth]{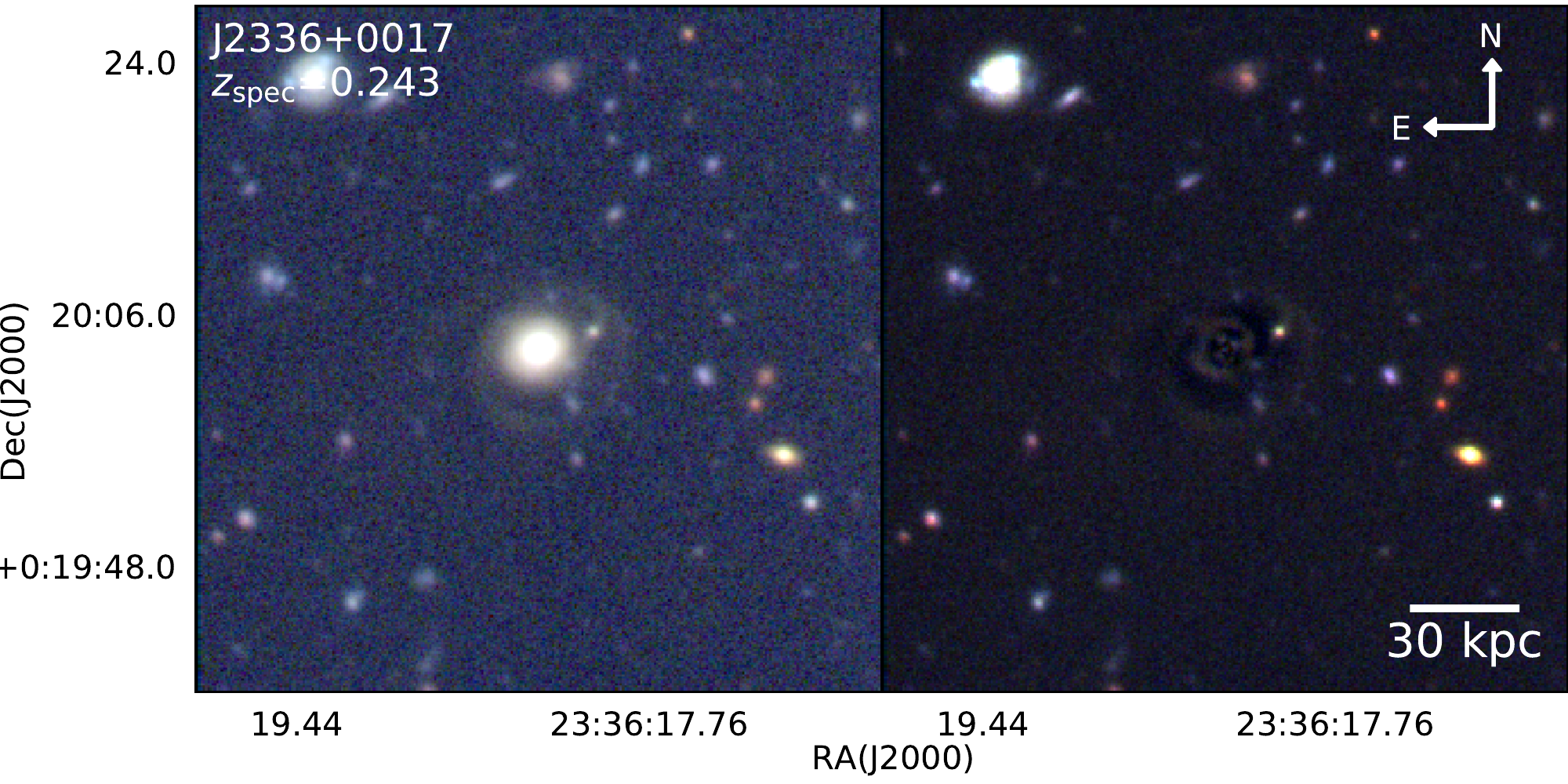}\includegraphics[width=0.4\textwidth]{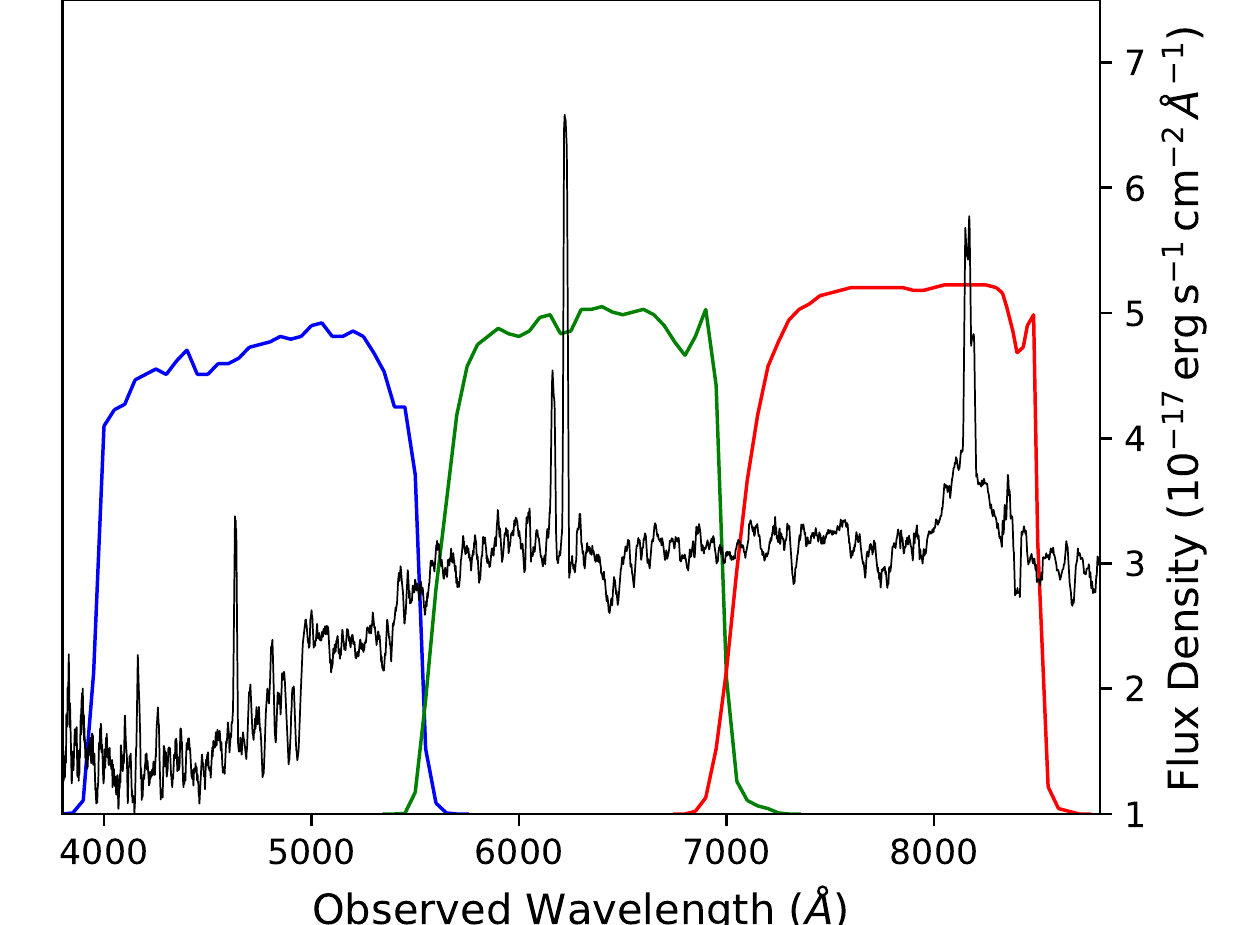}
	\caption{Left: 3-color images of each quasar host galaxy. Middle: 3-color residuals after model subtraction. Right: SDSS Spectra of each galaxy in its faint state. The GMOS broadband \textit{g}, \textit{r}, and \textit{i} bandpasses have been overlaid in blue, green, and red, respectively.}
    \label{fig:3_colour}
\end{figure*}

The model fit results for our quasar host galaxies are summarized in Tables \ref{tab:galfit_summ_1} \& \ref{tab:galfit_summ_2}. There is evidence for at least some disruption in each galaxy, primarily in the form of tidal streams. The exception to this is J0159+0033, which does not show an obvious tidal stream, but has faint extended regions to the Northwest and Southwest with diffuse emission in the \textit{r} and \textit{i} band residuals.

Our detailed fits prove effective at removing the continuum emission of each galaxy in all three bands and isolating any features that appear in only one or two bands. Ultimately we do not observe any voorwerpjes or EELR-like extended features. The single-band data, model, and residual images are available in Appendix \ref{sec:App_A}.

\begin{deluxetable*}{cccccccccc}
\tabletypesize{\footnotesize}
\tablecolumns{10}
\tablewidth{0pt}
\tablecaption{Summary of \textit{GALFIT} Results for Less Disturbed CL Quasar Hosts}
\tablehead{\colhead{Quasar Name} & \colhead{Band} & \colhead{PSF} & \colhead{Component} & \colhead{Component} & \colhead{Half-light Radius}  & \colhead{S\'ersic Index} & \colhead{Axis} & \colhead{Position} & \colhead{$\chi^2_{\nu}$} \vspace{-0.2cm}\\
    &  & \colhead{Magnitude} & \colhead{Label} & \colhead{Magnitude} & \colhead{$r_e$ (\asec)}  & \colhead{$n$} & \colhead{Ratio} & \colhead{Angle ($^\circ$)} &}
\startdata
    \ J0126$-$0839 & \textit{g} & $22.70\pm0.01$ & Single S\'ersic & $19.07\pm0.01$ & $0.721\pm0.002$ & $3.03\pm0.02$ & $0.93\pm0.01$ & $-60\pm3$ & $1.24$\\
	\ & & $21.54\pm0.01$ & S\'ersic 1 & $19.76\pm0.01$ & $0.552\pm0.004$ & $0.59\pm0.01$ & $0.84\pm0.01$ & $-62.0\pm0.5$ & $1.11$\\
    \ & & & S\'ersic 2 & $20.10\pm0.01$ & $2.31\pm0.01$ & $0.76\pm0.01$ & $0.89\pm0.01$ & $23.7\pm0.5$ & \\
    \ & \textit{r} & $22.95\pm0.01$ & Single S\'ersic & $18.31\pm0.01$ & $0.927\pm0.002$ & $3.60\pm0.01$ & $0.95\pm0.01$ & $-64.6\pm0.9$ & $1.40$\\
    \ & & $20.71\pm0.01$ & S\'ersic 1 & $19.18\pm0.01$ & $0.554\pm0.002$ & $0.67\pm0.01$ & $0.86\pm0.01$ & $-63.4\pm0.4$ & $1.18$\\
    \ & & & S\'ersic 2 & $19.25\pm0.01$ & $2.27\pm0.01$ & $0.77\pm0.01$ & $0.90\pm0.01$ & $28.8\pm0.4$ & \\
    \ & & & Gaussian & $23.29\pm0.01$ & $0.486\pm0.002$ & & $0.88\pm0.01$ & $12.8\pm0.4$ & \\
    \ & \textit{i} & $22.88\pm0.01$ & Single S\'ersic & $17.92\pm0.01$ & $0.823\pm0.002$ & $3.29\pm0.01$ & $0.94\pm0.01$ & $-70\pm1$ & 1.45\\
    \ & & $20.10\pm0.01$ & S\'ersic 1 & $18.83\pm0.01$ & $0.593\pm0.002$ & $0.56\pm0.01$ & $0.86\pm0.01$ & $-66.0\pm0.4$ & $1.20$\\
    \ & & & S\'ersic 2 & $18.87\pm0.01$ & $2.22\pm0.01$ & $0.79\pm0.01$ & $0.92\pm0.01$ & $32.9\pm0.4$ &\\
    \ & & & Gaussian & $21.77\pm0.01$ & $0.502\pm0.002$ & & $0.78\pm0.01$ & $2.5\pm0.4$ &\\
    \ J0159+0033 & \textit{g} & $23.81\pm0.01$ & Single S\'ersic & $20.12\pm0.01$ & $0.494\pm0.002$ & $2.52\pm0.02$ & $0.89\pm0.01$ & $60\pm1$ & $1.13$\\
    \ & & $22.89\pm0.06$ & S\'ersic 1 & $21.18\pm0.06$ & $2.0\pm0.1$ & $1.4\pm0.1$ & $0.80\pm0.01$ & $62\pm2$ & $1.11$\\
    \ & & & S\'ersic 2 & $20.58\pm0.06$ & $0.4\pm0.1$ & $0.8\pm0.1$ & $0.96\pm0.1$ & $73\pm2$ &\\
    \ & \textit{r} & $22.51\pm0.01$ & Single S\'ersic & $19.05\pm0.01$ & $0.500\pm0.002$ & $3.44\pm0.02$ & $0.89\pm0.01$ & $59.5\pm0.7$ & $1.16$\\
    \ & & $22.21\pm0.02$ & S\'ersic 1 & $20.66\pm0.02$ & $4.17\pm0.09$ & $0.71\pm0.03$ & $0.63\pm0.01$ & $52.4\pm0.8$ & $1.12$\\
    \ & & & S\'ersic 2 & $19.26\pm0.02$ & $0.40\pm0.09$ & $1.92\pm0.03$ & $0.91\pm0.01$ & $66.8\pm0.8$ &\\
    \ & \textit{i} & $23.27\pm0.01$ & Single S\'ersic & $18.57\pm0.01$ & $0.442\pm0.002$ & $3.62\pm0.01$ & $0.88\pm0.01$ & $57.5\pm0.4$ & $1.18$\\
    \ & & $21.15\pm0.05$ & S\'ersic 1 & $19.51\pm0.05$ & $1.83\pm0.09$ & $2.38\pm0.2$ & $0.80\pm0.01$ & $57.7\pm0.7$ & $1.11$\\
    \ & & & S\'ersic 2 & $19.30\pm0.05$ & $0.33\pm0.09$ & $1.3\pm0.2$ & $0.95\pm0.01$ & $70.9\pm0.7$ &\\
    \ J2336+0017 & \textit{g} & $23.16\pm0.01$ & Single S\'ersic & $20.44\pm0.01$ & $1.19\pm0.01$ & $2.02\pm0.03$ & $0.86\pm0.01$ & $-63\pm1$ & $1.12$\\
    \ & & $23.0\pm0.7$ & S\'ersic 1 & $21.9\pm0.7$ & $0.64\pm0.08$ & $1.3\pm0.2$ & $0.77\pm0.05$ & $-44\pm4$ & $1.12$\\
    \ & & & S\'ersic 2 & $20.8\pm0.7$ & $1.60\pm0.08$ & $1.5\pm0.2$ & $0.87\pm0.05$ & $-81\pm4$ &\\
    \ & \textit{r} & $21.92\pm0.01$ & Single S\'ersic & $19.39\pm0.01$ & $1.32\pm0.01$ & $1.87\pm0.01$ & $0.84\pm0.01$ & $-65.1\pm0.4$ & $1.24$\\
    \ & & $21.9\pm0.1$ & S\'ersic 1 & $22.0\pm0.1$ & $0.60\pm0.02$ & $1.4\pm0.1$ & $0.45\pm0.02$ & $-25.4\pm0.9$ & $1.23$\\
    \ & & & S\'ersic 2 & $19.5\pm0.1$ & $1.44\pm0.09$ & $1.6\pm0.1$ & $0.82\pm0.02$ & $-76.5\pm0.9$ &\\
    \ & \textit{i} & $21.25\pm0.01$ & Single S\'ersic & $18.96\pm0.01$ & $1.29\pm0.01$ & $1.75\pm0.01$ & $0.84\pm0.01$ & $-64.0\pm0.3$ & $1.24$\\
    \ & & $20.4\pm0.1$ & S\'ersic 1 & $21.3\pm0.1$ & $0.54\pm0.02$ & $1.7\pm0.1$ & $0.55\pm0.02$ & $-23.9\pm0.9$ & $1.22$\\
    \ & & & S\'ersic 2 & $19.0\pm0.1$ & $1.45\pm0.2$ & $1.5\pm0.1$ & $0.81\pm0.02$ & $-75.7\pm0.9$ &\\
									\enddata
\tablecomments{The single S\'ersic + PSF and double S\'ersic + PSF fits to the three CL quasar host galaxies well-described by a standard surface brightness profile. The PSF magnitude column contains the magnitude of the PSF component added to each type of fit. The position angle is defined with North $ = 0^\circ$ and increases as the long axis of the component rotates counter-clockwise. The Gaussian components added to the \textit{r} and \textit{i}-band  models of J0126$-$0839 are included in the fits due to their brightness impacting the reliability of our S\'ersic fits, but have not been subtracted from the residuals as we do not believe they are part of the galaxy's continuum emission. The goodness of fit, $\chi^2_\nu$, is given for the single S\'ersic + PSF and double S\'ersic + PSF fits on the corresponding rows.}
\label{tab:galfit_summ_1}
\end{deluxetable*}

\begin{deluxetable*}{cccccccccc}
\tabletypesize{\footnotesize}
\tablecolumns{10}
\tablewidth{0pt}
\tablecaption{Summary of \textit{GALFIT} Results for J1011+5442}
\tablehead{\colhead{Quasar Name} & \colhead{Band} & \colhead{PSF} & \colhead{Component} & \colhead{Component} & \colhead{$r_e$ (\asec)}  & \colhead{$n$} & \colhead{Axis} & \colhead{Position} & \colhead{$\chi^2_{\nu}$} \vspace{-0.2cm}\\
     &  & \colhead{Magnitude} & \colhead{Label} & \colhead{Magnitude} & \colhead{[$r_{\mathrm{break}}$, $l_{\mathrm{soft}}$]}  & \colhead{[$A_{1}$, $\phi_{1}$]} & \colhead{Ratio} & \colhead{Angle ($^\circ$)} &}
\startdata
    \ J1011+5442 & \textit{g} & $21.56\pm0.01$ & Single S\'ersic & $19.50\pm0.01$ & $1.44\pm0.01$ & $2.60\pm0.02$ & $0.94\pm0.01$ & $-63\pm1$ & $2.04$\\
    \ & & $22.37\pm0.02$ & S\'ersic 1 & $20.06\pm0.04$ & $2.0\pm0.3$ & $20\pm2$ & $0.34\pm0.01$ & $7.9\pm0.3$ & $1.30$\\
    \ & & & S\'ersic 2 (ext) & $20\pm200$ & $10\pm300$ & $0.02\pm0.01$ & $0.50\pm0.01$ & $87.5\pm0.3$ &\\
    \ & & & S\'ersic 3 (south) & $21.71\pm0.01$ & $2.45\pm0.01$ & $0.21\pm0.01$ & $0.24\pm0.01$ & $1.2\pm0.2$ &\\
    \ & & & S\'ersic-T & $23\pm1$ & $1.89\pm0.1$ & $0.59\pm0.01$ & $0.30\pm0.01$ & $-70.8\pm0.1$ &\\
    \ & & & Truncation & & $2\pm1$, $1\pm1$ & $0.9\pm0.1$, $50\pm20$ &  $0.5\pm0.1$ & $30\pm20$ &\\
    \ & \textit{r} & $21.50\pm0.01$ & Single S\'ersic & $18.66\pm0.01$ & $1.37\pm0.01$ & $3.38\pm0.02$ & $0.95\pm0.01$ & $21\pm1$ & $3.10$\\
    \ & & $21.71\pm0.01$ & S\'ersic 1 & $19.12\pm0.01$ & $3.45\pm0.07$ & $14.1\pm0.1$ & $0.43\pm0.01$ & $10.78\pm0.07$ & $1.38$\\
    \ & & & S\'ersic 2 (ext) & $20.81\pm0.01$ & $3.98\pm0.01$ & $0.02\pm0.01$ & $0.45\pm0.01$ & $88.4\pm0.1$ &\\
    \ & & & S\'ersic 3 (south) & $21.08\pm0.01$ & $2.44\pm0.01$ & $0.16\pm0.01$ & $0.25\pm0.01$ & $0.8\pm0.1$ &\\
    \ & & & S\'ersic-T & $22.0\pm0.7$ & $1.57\pm0.01$ & $0.78\pm0.01$ & $0.37\pm0.01$ & $-72.1\pm0.1$ &\\
    \ & & & Truncation & & $2.1\pm0.6$, $1.4\pm0.4$ & $0.80\pm0.08$, $50\pm10$ & $0.66\pm0.04$ & $40\pm10$ &\\
    \ & \textit{i} & $21.42\pm0.01$ & Single S\'ersic & $18.41\pm0.01$ & $1.36\pm0.01$ & $3.85\pm0.03$ & $0.91\pm0.01$ & $19.9\pm0.7$ & $2.38$\\
    \ & & $22.04\pm0.01$ & S\'ersic 1 & $18.80\pm0.01$ & $2.1\pm0.1$ & $13.1\pm0.4$ & $0.49\pm0.01$ & $10.6\pm0.1$ & $1.25$\\
    \ & & & S\'ersic 2 (ext) & $20\pm100$ & $10\pm200$ & $0.02\pm0.01$ & $0.45\pm0.01$ & $89.1\pm0.2$ &\\
    \ & & & S\'ersic 3 (south) & $20.92\pm0.01$ & $2.46\pm0.01$ & $0.16\pm0.01$ & $0.25\pm0.01$ & $0.4\pm0.1$ &\\
    \ & & & S\'ersic-T & $23.0\pm0.9$ & $1.65\pm0.01$ & $0.73\pm0.01$ & $0.36\pm0.01$ & $-70.7\pm0.1$ &\\
    \ & & & Truncation &   & $3.0\pm0.7$, $2.3\pm0.6$ & $0.82\pm0.06$, $51\pm3$ & $0.46\pm0.05$ & $24\pm5$ &\\
\enddata
\tablecomments{ Similar to Table \ref{tab:galfit_summ_1}, but for J1011+5442 which has a more disturbed host. The column headers in square brackets are specific to the truncated S\'ersic profile. $r_{\mathrm{break}}$ and $l_{\mathrm{soft}}$ are the break radius and softening length, respectively, and describe the annulus within which the S\'ersic profile fades from 0 to 100\% of its untruncated value. $A_1$ and $\phi_1$ describe the amplitude and phase angle of the first azimuthal Fourier mode component added to the truncation to allow it to deform (see the \textit{GALFIT} documentation for more information). In the multi-component fits, S\'ersic 1 is the central S\'ersic component. S\'ersic 2 models the emission from a patch offset $~30$kpc to the East present in all three bands. S\'ersic 3 models the emission extending south from the center of the galaxy, lying opposite the diffuse emission extending to the north. S\'ersic-T is a S\'ersic profile with an inner truncation and slight azimuthal twist representing the ring-like structure extending East of the bright core. These fits are noticeably worse than the less disturbed galaxies in our sample, and include extremely poorly constrained parameters in otherwise good fits (according to the reported $\chi^2_{\nu}$ values), or unrealistic parameters, such as S\'ersic indices of 20 or 0.02. For these reasons we exclude J1011+5442 from further analysis when describing overall properties of the shapes of our CL quasar host galaxies, elsewhere in this work.}
\label{tab:galfit_summ_2}
\end{deluxetable*}

\subsection{J0126$-$0839}
\label{subsec:J0126}
This galaxy appears to be surrounded by a shell-like structure, which could be indicative of past mergers \citep{1992ApJ...399L.117H}. This galaxy appears to have a large patch of diffuse emission to its Northwest (circled in Figure \ref{fig:3_colour}). This diffuse emission region is spatially separated from the bulge and disk of the host galaxy, but appears to form the end of a large tidal tail connected to the galaxy in the residuals of all three bands. Another notable feature of this galaxy is a bright spot east of the nucleus in the \textit{r} and \textit{i} band images, easily visible in the residuals. The most prominent emission lines in the \textit{r} and \textit{i} bands are from [O III] and H$\alpha$, respectively. While this does not appear to be a voorwerp, as its H$\alpha$ emission is too strong, it may be the nucleus of the other galaxy involved in its most recent merger. The wavelength of the lines identified as H$\alpha$ do not indicate that this is a background quasar, and furthermore we do not see the broad Balmer emission from a background quasar in the optical spectrum of this object 
\subsection{J0159+0033}
\label{subsec:J0159}
J0159+0033 does not show the dramatic features (tidal tails, shells) indicating recent mergers that the other galaxies in our sample do. The overall shape of the galaxy appears slightly irregular, concave toward the Southeast. Subtraction of the galaxy's continuum reveals faint emission extending $\sim$20 kpc to the Southwest in \textit{r} and \textit{i} (circled in Figure \ref{fig:3_colour}) that is responsible for its apparent shape. There is some evidence for this feature in the \textit{g} residuals, but it is noticeably fainter. Given that the \textit{g} image contains one more exposure than \textit{r} this is unlikely to be an effect of differing depths, and, while not a voorwerp, may be evidence for tidal debris.

\subsection{J1011+5442}
\label{subsec:J1011}
This is this is the most obviously irregular galaxy in our sample. It is evident that this galaxy has recently undergone a merger that has resulted in a significantly disrupted galaxy. The tidal streams (regions B and D in Figure \ref{fig:3_colour}) indicate that the planes of motion of each galaxy in the merger were likely perpendicular to each other. While the [O III] emission of this galaxy is significant, the tidal features and disruption make it difficult to distinguish any voorwerpjes or EELR surrounding it.

To fit the stellar light from the galaxy, more components were required than the double-S\'ersic profile employed for the other galaxies of our sample. The model consists of: (1) a central S\'ersic component, (2) an offset and elongated S\'ersic component to model the emission in the South (region c) which sits opposite the long diffuse tidal tail in the north (region A), (3) a S\'ersic profile with (4) an inner truncation to represent the ring-like East-West tidal feature, and (5) another S\'ersic profile offset to the East to model emission extending from the tidal stream (region c). 

\subsection{J2336+0017}
\label{subsec:J2336}
Similar to J0126$-$0839, this galaxy shows evidence for large-scale tidal streams, or shells in all three bands. In this case, there is a significant difference in depth between the \textit{g} band image and the \textit{r} and \textit{i} band images. Nevertheless, there are no obvious features which appear in \textit{r} that are strong candidates for extended emission line regions. 

\section{Discussion}
\label{sec:disc}

\subsection{Voorwerpjes}
\label{subsec:voorwerps}
Voorwerpjes are clouds of gas that glow primarily in [O III]. In order for voorwerpjes or EELRs to be visible, we require a sufficient quantity of gas to be present $>$10 kpc from the nucleus, at such a density that the upper level of the transition not de-excited collisionally. Many EELR/voorwerpjes appear to have a biconical structure that originates from the nucleus (\citealt{2015AJ....149..155K} and \citealt{2018MNRAS.480.2302S} have high resolution examples of such morphologies from \textit{HST} and Subaru, respectively). These observations initially led to the assumption that the gas in these regions was due to outflows from quasar winds. Using IFU spectroscopy to study the gas kinematics has shown that this is only true in radio-loud quasars \citep{2006NewAR..50..694S, 2009ApJ...690..953F}. Near radio-quiet quasars, the biconical structure of EELRs generally seems to be the result of gas in orbit around the host galaxy, likely from tidal debris \citep{2013A&A...549A..43H, 2013ApJ...773..148S, 2014ApJ...792...72G, 2015AJ....149..155K, 2017ApJ...835..256K}.

Broad-band imaging has been used to detect EELRs and voorwerps around faded quasars \citep{2018MNRAS.480.2302S}, including the prototypical Hanny's Voorwerp \citep{2012AJ....144...66K}. In these cases, the [O III] emitting regions are visually distinct from the underlying galaxy. The bandpasses that were used for our science images were such that the voorwerpjes (if present) should appear obvious in the \textit{r}-band imaging. \cite{2012AJ....144...66K} show that Hanny's Voorwerp emits H$\alpha$ using narrow-band imaging, which should be observed in the \textit{i}-band in all of our sample quasars except J0159+0033. These distinct features are absent from our \textit{r}-band images when compared to the images in bands free of strong emission lines, and there is no trace of the characteristic green emission present in the \textit{gri} composites or the residuals shown in Figure \ref{fig:3_colour}.

From our residuals it is evident that three of our four CL quasar host galaxies have undergone recent tidal interactions capable of placing gas where it can be ionized by the quasar. The absence of EELR or voorwerpjes in our sample may simply be a matter of chance. A multi-band search conducted by \cite{2012JSARA...5...29C} of around 30 AGN found that an EELR was seen in only half of their sample, of which two thirds showed significant disruption. Their galaxies were selected for having voorwerp-like features in the band containing [O III] emission, and as such we should expect them to be biased toward finding voorwerpjes. When we consider that of the entire population of galaxies in their original Galaxy Zoo sample ($\sim10^6$) only 15 were found to contain voorwerpjes, we should be unsurprised that our CL quasar host galaxies lack them.

\subsection{Quasar Host Galaxy Photometric Properties}

Fitting the continuum emission of our CL quasar host galaxies offers important morphological information that allows for a comparison to other quasar hosts and inactive galaxies. Due to the significant evolution of rest-frame galaxy colors with redshift, we will limit our comparison to quasar host galaxy studies close to the redshift range of our sample (\textit{z}$\sim$0.2-0.3).

Rest frame $g-i$ colors and absolute magnitudes were determined for our CL quasar host galaxies using apparent magnitudes from our overall S\'ersic fits. The k-corrections were calculated using a \textit{Python} implementation\footnote{https://github.com/nirinA/kcorrect\_python} of \textit{kcorrect} version 4.3\footnote{http://kcorrect.org/} \citep{2007AJ....133..734B} and the colors were corrected for dust reddening using the maps of \cite{1998ApJ...500..525S}. This procedure and the results are summarized in Table \ref{tab:c_mag_table}.

\begin{figure}
	\includegraphics[width=\columnwidth]{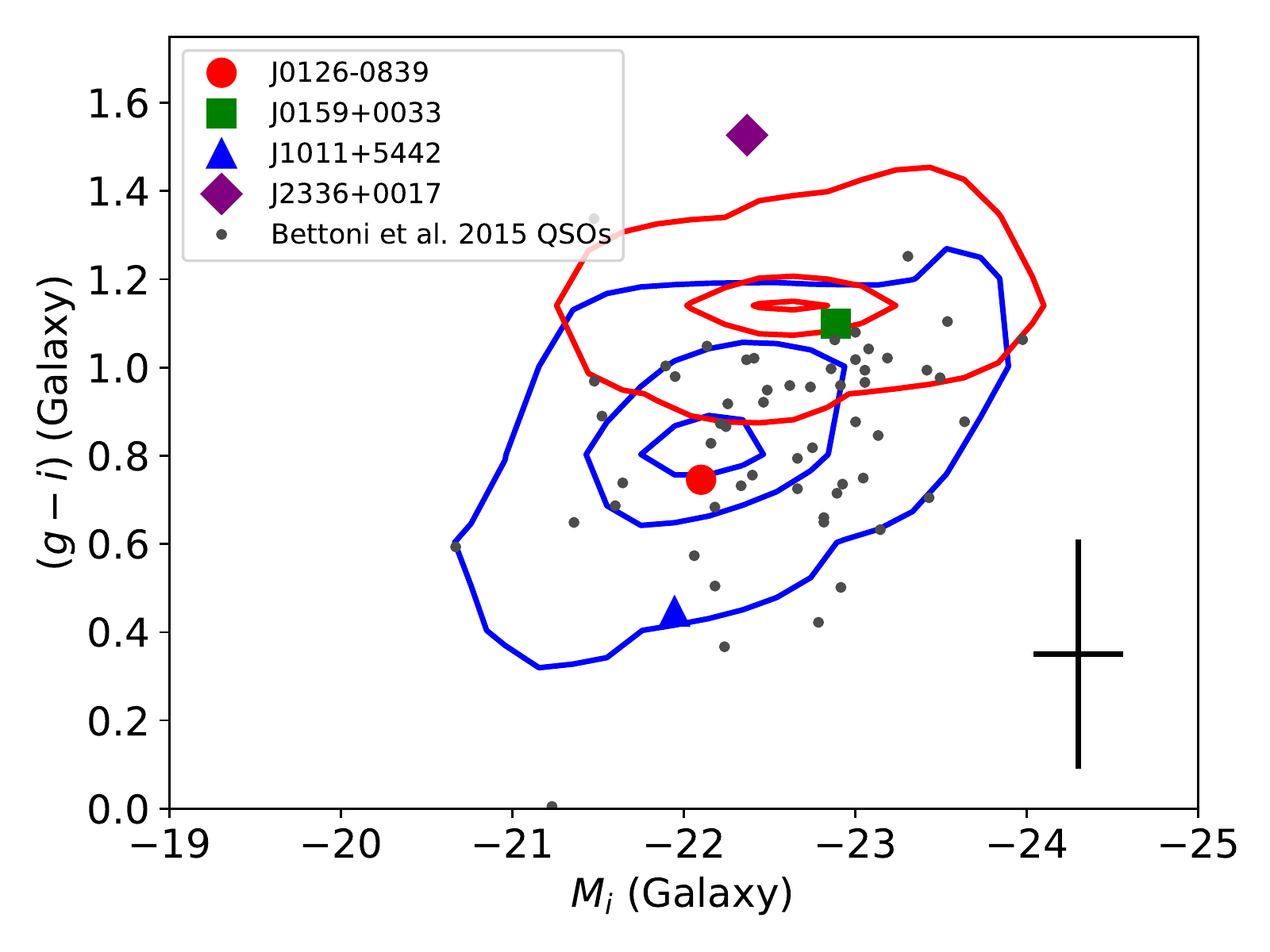}
    \caption{Quasar host galaxy rest frame colors and \textit{i}-band magnitudes. The contours are the CMD of inactive galaxies from the MPA-JHU catalog between $0.1 \leqslant z \leqslant 0.6$ which have been split into the red sequence and blue cloud (red and blue contours respectively, containing 90\%, 50\%, and 20\% of each subsample) using the $u-r$ color split prescribed by \cite{2001AJ....122.1861S}. The CL quasar host galaxies have green colors on average, and sit in the blue cloud with other AGN host galaxies. The small dots represent the quasar host galaxies from \cite{2015MNRAS.454.4103B}. Representative error bars based on \cite{2015MNRAS.454.4103B} have been included.}
    \label{fig:colour_mag}
\end{figure}

Studies of quasar host galaxies have found them to be relatively blue and high mass ($M_{*} \geqslant 10^{10} M_{\odot}$) when compared to inactive galaxies (\citealt{2003MNRAS.346.1055K, 2013ApJ...763..133T, 2014ApJ...780..162M, 2015MNRAS.454.4103B}), with no significant differences between Type 1 and Type 2 quasar hosts. We compare the color-magnitude diagram (CMD) of our CL quasar host galaxies to inactive galaxies found in the MPA-JHU DR7 catalog\footnote{https://wwwmpa.mpa-garching.mpg.de/SDSS/DR7/} and the quasar host galaxies of \cite{2015MNRAS.454.4103B} in Figure \ref{fig:colour_mag}. For the inactive galaxies we used recent model magnitudes from SDSS DR14 \citep{2018ApJS..235...42A} to directly compare with our host galaxy magnitudes. The cModel magnitudes we used were derived from the linear combination of the best-fit exponential and De Vaucoleurs profile fits. These are then dereddened and k-corrected. 

We find that our CL quasar host galaxies do not appear to lie on the red sequence, instead sitting within the blue cloud like most star-forming and AGN host galaxies. J2336+0017 (indicated by the purple diamond in Figure \ref{fig:colour_mag}) sits above the red sequence in the rest frame, but this is not unheard of for quasar host galaxies. \cite{2014ApJ...780..162M} find several hosts that are redder than inactive galaxies on the red sequence, with no trend in quasar luminosity. We perform a 2-sample KS test in the color-magnitude plane comparing the CL quasar host galaxies to the red sequence and blue cloud galaxies of Figure \ref{fig:colour_mag}. We find that our CL quasar host galaxies are more likely to have been drawn from the blue cloud than from the red sequence, with p-values of $\sim0.2$ and $\sim0.08$, respectively.

From our analysis of the host galaxy colors, we can conclude that CL quasar host galaxies are drawn from a similar population as Type 1 and Type 2 quasar hosts. These appear to be galaxies that still have a significant amount of ongoing star formation. 

\begin{deluxetable*}{cccccccccc}
\tabletypesize{\footnotesize}
\tablecolumns{10}
\tablewidth{0pt}
\tablecaption{CL Quasar Host Galaxy K-Corrections and Colors}
\tablehead{\colhead{Quasar Name} & \colhead{$z_{\mathrm{spec}}$} & \colhead{$m_g$} & \colhead{$m_i$} & \colhead{E(B-V)} &\colhead{$k_g$} & \colhead{$k_i$} & \colhead{$M_g$} & \colhead{$M_i$} &\colhead{\textit{g}-\textit{i} Color}}
\startdata
J0126$-$0839 & 0.198 & $19.07\pm0.01$ & $17.92\pm0.01$ & 0.02984 & 0.380 & 0.027 & $-21.36\pm0.01$ & $-22.10\pm0.01$ & $0.75\pm0.02$\\
J0159+0033 & 0.312 & $20.12\pm0.01$ & $18.57\pm0.01$ & 0.02634 & 0.754 & 0.348 & $-21.79\pm0.01$ & $-22.89\pm0.01$ & $1.10\pm0.02$\\
J1011+5442 & 0.246 & $19.50\pm0.01$ & $18.41\pm0.01$ & 0.08757 & 0.503 & -0.125 & $-21.50\pm0.01$ & $-21.95\pm0.01$ & $0.45\pm0.02$\\
J2336+0017 & 0.243 & $20.44\pm0.01$ & $18.96\pm0.01$ & 0.03062 & 0.734 & 0.443 & $-20.84\pm0.01$ & $-22.37\pm0.01$ & $1.53\pm0.02$\\
\enddata
\tablecomments{We use our single-S\'ersic fit magnitudes ($m_g$, $m_i$) along with calculated extinction values ,E(B-V), and k-corrections ($k_g$, $k_i$) to determined the rest frame absolute magnitudes in \textit{g} and \textit{i} ($M_g$, $M_i$) by finding the distance modulus and k \& extinction correcting.}
\label{tab:c_mag_table}
\end{deluxetable*}

\subsection{Quasar Host Galaxy Morphology and Dynamics}
Classifying the morphology of CL quasar host galaxies can provide information about their evolutionary history, or illustrate why they may differ from standard quasars. CL quasar host galaxy morphologies have not been examined previously, partially due to the low resolution of available imaging, but primarily due to the QSO outshining the galaxy in its bright state. The images obtained for this work allow for detailed multi-component modeling used to search for voorwerpjes and allow for reliable fits to the overall surface brightness profiles of the host galaxies provided that they are not too disturbed. Through these fits we can classify each galaxy's overall morphology.

Of our galaxies well fit by S\'ersic profiles (all except J1011+5442), two of three have $\left\langle n \right\rangle > 3$, with the remaining having $\left\langle n \right\rangle = 1.88$ when averaged across all three imaging bands. Based on the intermediate $n$ values between those of disks and ellipticals, we classify our CL quasar hosts as spheroidal. Inspection of the images and residuals show that the CL quasar host galaxies lack strong spiral features, but also lack the very extended profiles expected from elliptical galaxies, supporting this conclusion.

The primary caveat to our multi-band analysis is that different bands sample different stellar populations, and unless the various populations are well-mixed, we should expect a wavelength-dependent S\'ersic index. \cite{2014MNRAS.441.1340V} sort galaxies into disk-like and elliptical based on their S\'ersic index in a fiducial band; they find that disk-like galaxies tend to have larger S\'ersic indices at longer wavelengths, while elliptical galaxies have relatively constant S\'ersic indices. \cite{2015MNRAS.454..806K} extend this analysis and find that more luminous disks show a stronger wavelength dependent S\'ersic index. Keeping this in mind, we examine several works below as a comparison sample. 
%A significant amount of work has been done previously to classify quasar host galaxies. These have generally been biased toward studying Type 2 QSOs due to their fainter nuclear emission, or low redshift Type 1 QSOs as they are more easily resolved. We have the ability to make these classifications for galaxies that recently hosted bright AGN. However, we must take care before drawing conclusions. The primary caveat to our multi-band analysis is that different bands sample different stellar populations, and unless the various populations are well-mixed, we should expect a wavelength-dependent S\'ersic index. \cite{2014MNRAS.441.1340V} sort galaxies into disk-like and elliptical based on their S\'ersic index in a fiducial band; they find that disk-like galaxies tend to have larger S\'ersic indices at longer wavelengths, while elliptical galaxies have relatively constant S\'ersic indecies. \cite{2015MNRAS.454..806K} extends this analysis and find that more luminous disks show a stronger wavelength dependent S\'ersic index. Keeping this in mind, we examine several works below as a comparison sample.

\cite{2003MNRAS.346.1055K} primarily studied type 2 Seyfert galaxies and more distant type 2 quasars but include a comparison with broad line quasars similar to our CL quasar sample in their bright states. Morphologically, they classify galaxies as disk-like or elliptical using the SDSS concentration parameter $C$ (\citealt{2001AJ....122.1238S,2001AJ....122.1861S}). They find that AGN are slightly biased toward higher concentrations, preferring elliptical, spheroidal, and disky-elliptical hosts, when compared to the general population of non-AGN emission-line galaxies which prefer disks. In order to compare with our fitted S\'ersic indices, we use Table 1b of \cite{2005AJ....130.1535G} to interpolate concentration values for our measured S\'ersic indices. The result is summarized in Table \ref{tab:sersic_conc}. We find that our quasar host galaxies have concentration indices $> 2.7$, consistent with the majority of \cite{2003MNRAS.346.1055K} AGN, as compared to the general star-forming galaxy population, which, on average prefer a lower concentration index of $\sim2.1$.

\begin{deluxetable}{cccc} 
\tabletypesize{\footnotesize} 
\tablecolumns{5} 
\tablewidth{0pt} 
\tablecaption{S\'ersic Index to SDSS Concentration Index} 
\tablehead{\colhead{Quasar Name} & \colhead{Band} & \colhead{S\'ersic $n$} & \colhead{Concentration $C_{R90/R50}$}}
\startdata
J0126$-$0839 & \textit{g} & 3.03 & 3.18\\
 & \textit{r} & 3.44 & 3.26\\
 & \textit{i} & 3.29 & 3.23\\
J0159+0033 & \textit{g} & 2.52 & 3.05\\
 & \textit{r} & 3.44 & 3.26\\
 & \textit{i} & 3.62 & 3.29\\
J2336+0017 & \textit{g} & 2.02 & 2.89\\
 & \textit{r} & 1.87 & 2.83\\
 & \textit{i} & 1.75 & 2.77\\
\enddata 
%\vspace{-0.8cm} 
\tablecomments{Conversion of \textit{GALFIT}-derived S\'ersic indices to SDSS concentration index for our less-disrupted CL quasar host galaxies. Our quasar host galaxies prefer concentrations $>2.7$, which classifies them as spheroidal or elliptical.}
\label{tab:sersic_conc}
\end{deluxetable}

The sample of \cite{2003MNRAS.340.1095D} focuses on radio quiet and radio loud quasars that have matched \textit{Hubble Space Telescope} \textit{V}-band luminosities, and redshifts similar to our CL quasar hosts, and are composed of a mixture of type 1, type 2, and intermediate AGN. Our quasars lie in the VLA FIRST survey coverage and have no current radio detections, hence we consider them to be radio quiet. We compare with this sample as a way of determining whether diffraction-limited imaging may give different morphology information due to its ability to better resolve details in the centers of galaxies. In comparing the goodness-of-fit between an exponential disk, and a De Vaucouleurs profile to their quasar host galaxies, \cite{2003MNRAS.340.1095D} also find that their quasar sample is better-fit by the more extended De Vaucouleurs profile, and so are near-universally spheroidal. We perform a similar analysis by using \textit{GALFIT} to fit a single exponential disk or De Vaucouleurs component to our quasar host galaxies. The results of this analysis are contained in Table \ref{tab:devauc_exp}. We also find that our quasar host galaxies are better modeled by the more spheroidal De Vaucoleurs profile, as they provide a $\chi^{2}$ closer to our multi-component fits in Table \ref{tab:galfit_summ_1}. 

\cite{2003MNRAS.340.1095D} also find that the galaxies that show a significant disk component are the least luminous of the radio quiet quasar sample, and infer a relationship between the presence of a disk and low optical luminosity. Though our sample is small, of our three galaxies well described by S\'ersic profiles, J2336+0017 which has $\left\langle n \right\rangle = 1.88$ is approximately 0.5 dex fainter in its bright state than J0126$-$0839 and J0159+0033, which have $\left\langle n \right\rangle \approx 3.2$. If we take a smaller S\'ersic index to imply that a galaxy is more disk-like, then we see a relationship between morphology and (bright state) quasar luminosity consistent with the findings of \cite{2003MNRAS.340.1095D}.

\begin{deluxetable}{ccccc} 
\tabletypesize{\footnotesize} 
\tablecolumns{5} 
\tablewidth{0pt} 
\tablecaption{Exponential vs De Vaucouleurs Fit Comparison} 
\tablehead{\colhead{Quasar Name} & \colhead{Band} & \colhead{$\chi^2_{\mathrm{Exponential}}$} & \colhead{$\chi^2_{\mathrm{De Vaucouleurs}}$} & \colhead{Best Fit} }
\startdata
J0126$-$0839 & \textit{g} & 1.842 & 1.279 & De Vauc\\
 & \textit{r} & 4.226 & 1.465 & De Vauc\\
 & \textit{i} & 4.185 & 1.518 & De Vauc\\
J0159+0033 & \textit{g} & 1.233 & 1.168 & De Vauc\\
 & \textit{r} & 1.758 & 1.185 & De Vauc\\
 & \textit{i} & 2.746 & 1.215 & De Vauc\\
J2336+0017 & \textit{g} & 1.298 & 1.133 & De Vauc\\
 & \textit{r} & 2.743 & 1.335 & De Vauc\\
 & \textit{i} & 4.361 & 1.515 & De Vauc\\
\enddata 
\tablecomments{A direct comparison with the analysis done by \cite{2003MNRAS.340.1095D}. Our CL quasar host galaxies are better fit by De Vaucouleurs surface brightness profiles than exponential profiles, consistent with their findings.}
\label{tab:devauc_exp}
\end{deluxetable}

\begin{figure}
	\includegraphics[width=\columnwidth]{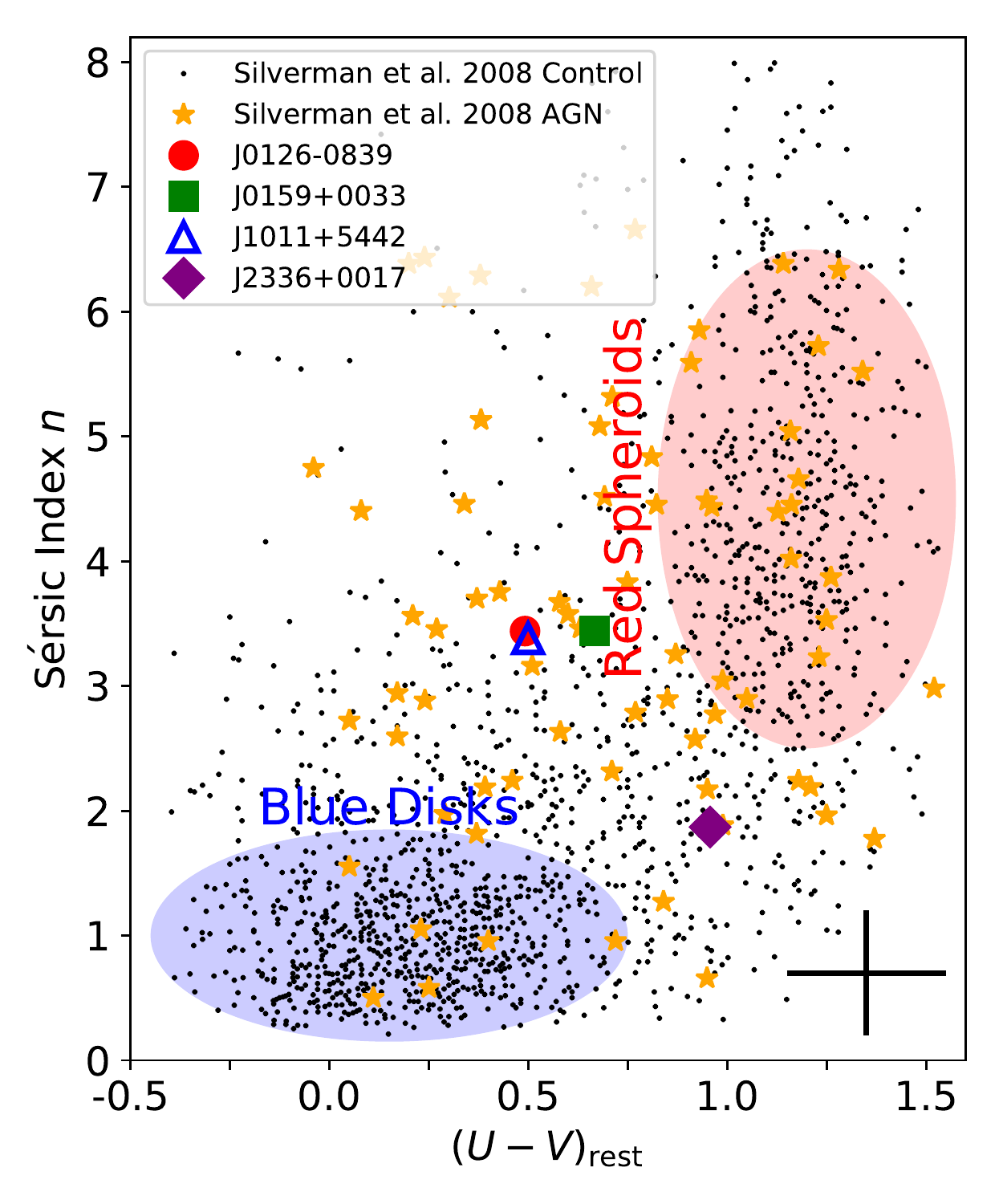}
    \caption{The colors and S\'ersic indices of our changing look quasar host galaxies compared with \cite{2008ApJ...675.1025S} AGN host galaxies and inactive galaxies (see their Figure 10). We see that our CL quasar hosts lie in the "green valley", i.e. they are redder and more spheroidal than inactive blue galaxies, like the majority of AGN. We use the \textit{r} band S\'ersic indices for our host galaxies, as it lies the closest to the wavelength used for the morphology measurements in the comparison sample. Representative error bars based on \cite{2008ApJ...675.1025S} are included.}
    \label{fig:colour_sersic}
\end{figure}

In combining colors and the S\'ersic index of galaxies we can see that AGN hosts are distinct from both inactive ellipticals and star-forming disks. In Figure \ref{fig:colour_sersic} we plot our CL quasar host galaxies (that are well-fit by S\'ersic profiles) in the rest frame $(U-V)$ color-S\'ersic index plane along with the sample of luminous ($M_V < -20.7$) AGN and inactive galaxies of \cite{2008ApJ...675.1025S}. The disk galaxies are distinctly blue and have low S\'ersic indices, while ellipticals are red with higher S\'ersic indices. AGN hosts are scattered, but predominantly reside in the "green valley" a relatively sparse region between the two populations. It is in this region that our three CL quasar host galaxies reside, consistent with other AGN.  

The relation between mergers and AGN activity has been debated considerably. It is clear from the residuals that the CL quasar host galaxies all show some measure of disruption (See Figure \ref{fig:3_colour} or Figures \ref{fig:J0126_and_J0159} \& \ref{fig:J1011_and_J2336} for single-band images). Though determinations of the mass ratio of the merger progenitors is beyond the scope of this work, at least one (J1011+5442) is likely to have been a merger between two galaxies of comparable mass by inspection of its environment, alone. There is evidence that the presence of tidal tails indicates a merger of mass ratio of at least 1/3 within the last $\sim$Gyr \citep{1995ApJ...438L..75M,2000RSPTA.358.2063S,2009ApJ...697.1971J}. \cite{2003MNRAS.340.1095D} note that the radio quiet quasars in their sample also show tidal features that the radio loud quasars lack, indicating that radio observations of CL quasars may provide useful information regarding their dynamical history. 

A potential link between merger activity and the changing look phenomenon is observed in the CL Quasar, Mrk 1018, which has transitioned between spectral types twice in its observed history \citep{2016A&A...593L...8M,2017A&A...607L...9K}. The mechanism for the changes proposed by \cite{2018ApJ...861...51K} is that a recoiling SMBH is responsible for perturbing the accretion flow on approximately a 29 year period, corresponding to spectral changes in the quasar. They predict that this will lead to another change in the spectral type in the mid-2020s.

% ====================================
\section{Conclusions}
\label{sec:summ}
We investigated a sample of four faded CL quasars using broadband GMOS imaging to search for voorwerpjes ionized by the quasar in its bright state. We fit their stellar continuum images to reveal faint features, but did not detect voorwerpjes around our CL quasar host galaxies.

The now-faded quasars also allow us to characterize the morphology and photometry of their host galaxies. In a color-magnitude diagram we find that the CL quasar hosts lie in the blue cloud along with star forming galaxies and other typical AGN hosts, rather than in the red sequence with quiescent galaxies. Classifying them by their S\'ersic indcies adds to this distinction. In a color-S\'ersic index plot, the CL quasar hosts lie in the "green valley" between the two main populations, red elliptical galaxies, characterized by a De Vaucouleurs surface brightness profile, and star-forming disk galaxies characterized by an exponential profile. This region is favored by other AGN host galaxies, indicating that, in regards to their stellar population and overall morphology, CL quasars are similar to other AGN hosts.

Detailed subtraction of the surface brightness profiles of our CL quasar hosts reveals faint, secondary, morphological features present in all three imaging bands. We observe strong evidence for recent or ongoing mergers in three of our four CL quasar host galaxies. The primary indicators of these interactions are tidal tails observed in J0126$-$0839 and J2336+0017, and the bright tidal streams indicating two merging galaxies in J1011+5442. There is also a hint of diffuse emission and a slight irregularity in the shape of J0159+0033, however, these features remain faint even when the stellar continuum is subtracted, and thus are difficult to characterize in detail with the available imaging.

Our investigation has revealed that faded CL quasars are hosted in galaxies very similar to the population of other Type 1 and Type 2 quasars. In this sample we do not detect obvious voorwerpjes. We conclude that CL quasars hosts are typical AGN hosts that appear to be undergoing mergers or interactions with nearby galaxies. Due to the general scarcity of voorwerpjes, this analysis should be extended to a larger sample of CL quasars to improve the chances of detecting one, as voorwerpjes provide a powerful way of analyzing a quasar's luminosity history.
\\
\par
P.C. would like to acknowledge the helpful correspondence with Michael R. Blanton and Yoshiki Matsuoka regarding \textit{kcorrect}. P.C. would also like to acknowledge the Gemini staff for their assistance with questions regarding image data reduction and the observations. 

P.C., J.J.R., and D.H. acknowledge support from a Natural Sciences and Engineering Research Council of Canada (NSERC) Discovery grant and a Fonds de recherche du Québec–Nature et Technologies (FRQNT) Nouveaux Chercheurs grant. J.J.R. acknowledges funding from the McGill Trottier Chair in Astrophysics and Cosmology. D.H. acknowledges support from the Canadian Institute for Advanced Research (CIFAR).

Funding for the Sloan Digital Sky Survey IV has been provided by the Alfred P. Sloan Foundation, the U.S. Department of Energy Office of Science, and the Participating Institutions. SDSS acknowledges support and resources from the Center for High-Performance Computing at the University of Utah. The SDSS web site is www.sdss.org.

SDSS is managed by the Astrophysical Research Consortium for the Participating Institutions of the SDSS Collaboration including the Brazilian Participation Group, the Carnegie Institution for Science, Carnegie Mellon University, the Chilean Participation Group, the French Participation Group, Harvard-Smithsonian Center for Astrophysics, Instituto de Astrof\'isica de Canarias, The Johns Hopkins University, Kavli Institute for the Physics and Mathematics of the Universe (IPMU) / University of Tokyo, the Korean Participation Group, Lawrence Berkeley National Laboratory, Leibniz Institut f\"ur Astrophysik Potsdam (AIP), Max-Planck-Institut f\"ur Astronomie (MPIA Heidelberg), Max-Planck-Institut f\"ur Astrophysik (MPA Garching), Max-Planck-Institut f\"ur Extraterrestrische Physik (MPE), National Astronomical Observatories of China, New Mexico State University, New York University, University of Notre
Dame, Observat\'orio Nacional / MCTI, The Ohio State University, Pennsylvania State University, Shanghai Astronomical Observatory, United Kingdom Participation Group, Universidad Nacional Aut\'onoma de M\'exico, University of Arizona, University of Colorado Boulder, University of Oxford, University of Portsmouth, University of Utah, University of Virginia, University of Washington, University of Wisconsin, Vanderbilt University, and Yale University.

\software{SExtractor \citep{1996A&AS..117..393B}, PyRAF \citep{2012ascl.soft07011S}, GALFIT3 \citep{2002AJ....124..266P, 2010AJ....139.2097P}, GALFITM \citep{MegaMorph}, kcorrect (v4.34; \citealt{2007AJ....133..734B}), astropy \citep{2013A&A...558A..33A, 2018AJ....156..123A}}

\bibliographystyle{aasjournal}
\bibliography{main}

\appendix
\section{Galfit Input, Model, and Output Images}
\label{sec:App_A}
This appendix contains the input cutouts, model images, and residuals generated by \textit{galfit} in each band for the four changing look quasar host galaxies investigated in this work. The noise that appears in the model panels is due to our choice to use empirical PSF cutouts, rather than modeled PSFs. In our testing, this choice did not diminish our ability to find extended emission features in the residuals, and prevented significant oversubtraction of the galactic nucleus. 

\begin{figure}
	\includegraphics[width=\textwidth,trim={0 1.81cm 0 1.8cm},clip]{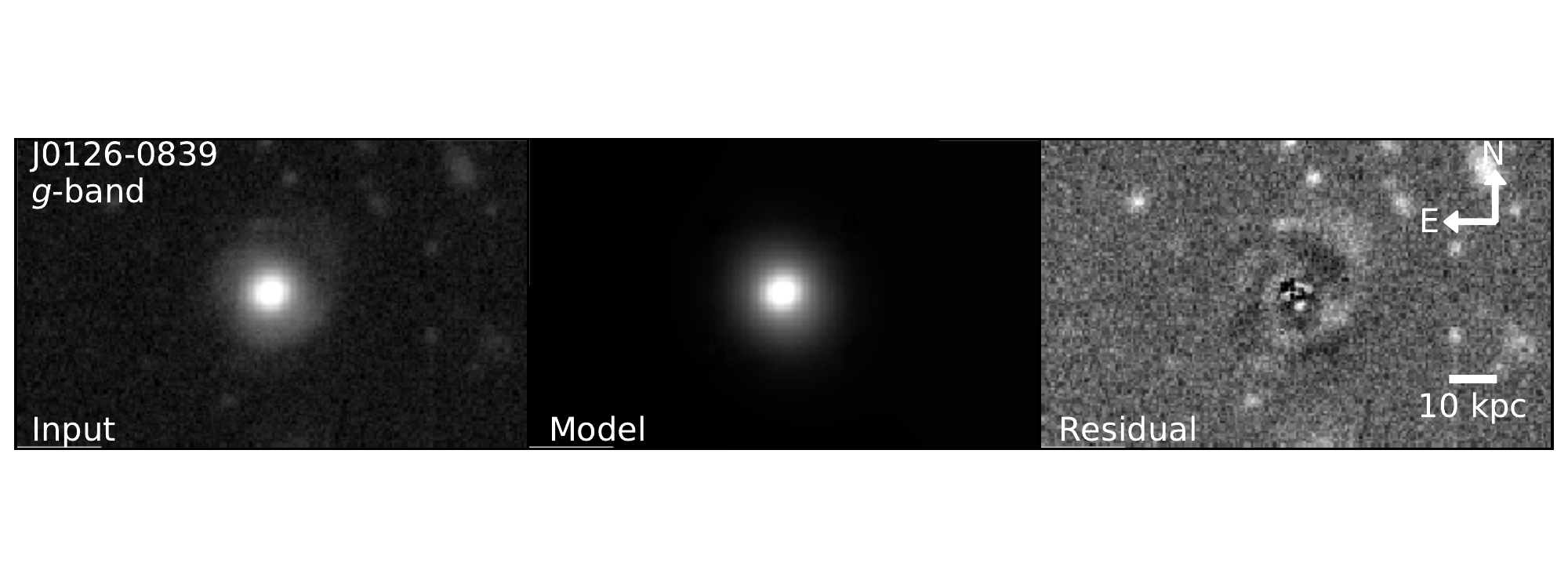}
    \includegraphics[width=\textwidth,trim={0 1.81cm 0 1.8cm},clip]{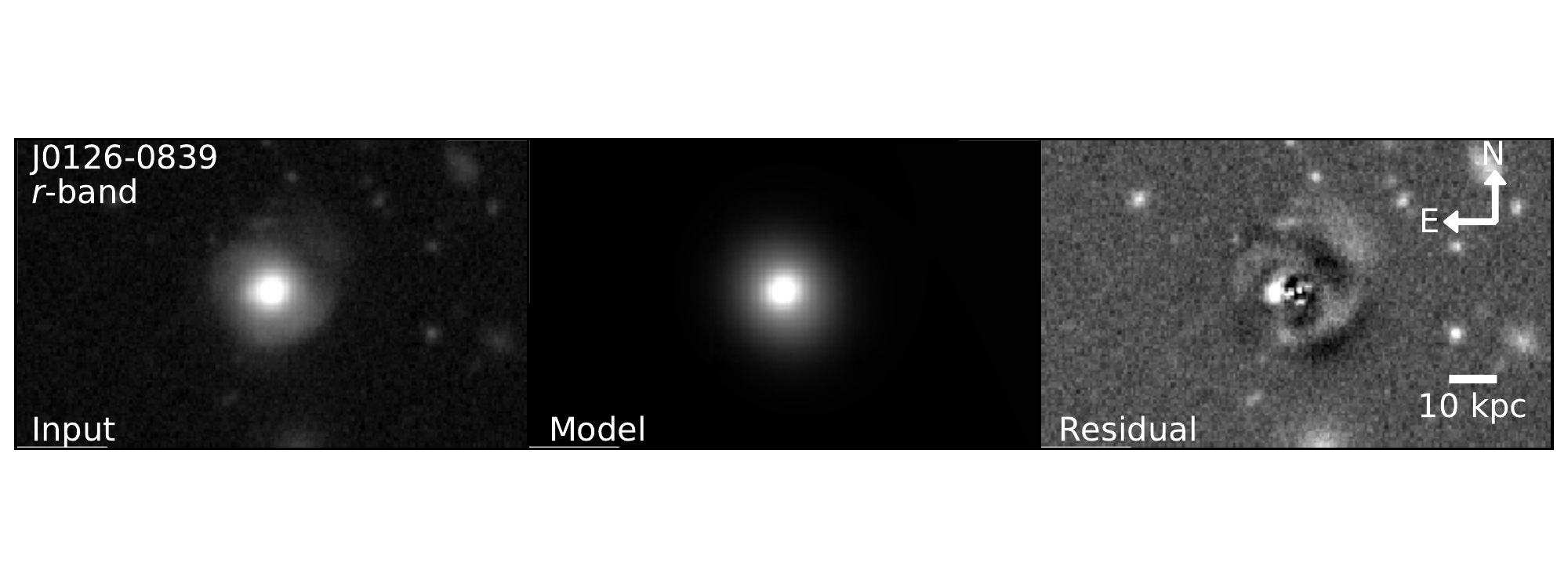}
    \includegraphics[width=\textwidth,trim={0 1.6cm 0 1.8cm},clip]{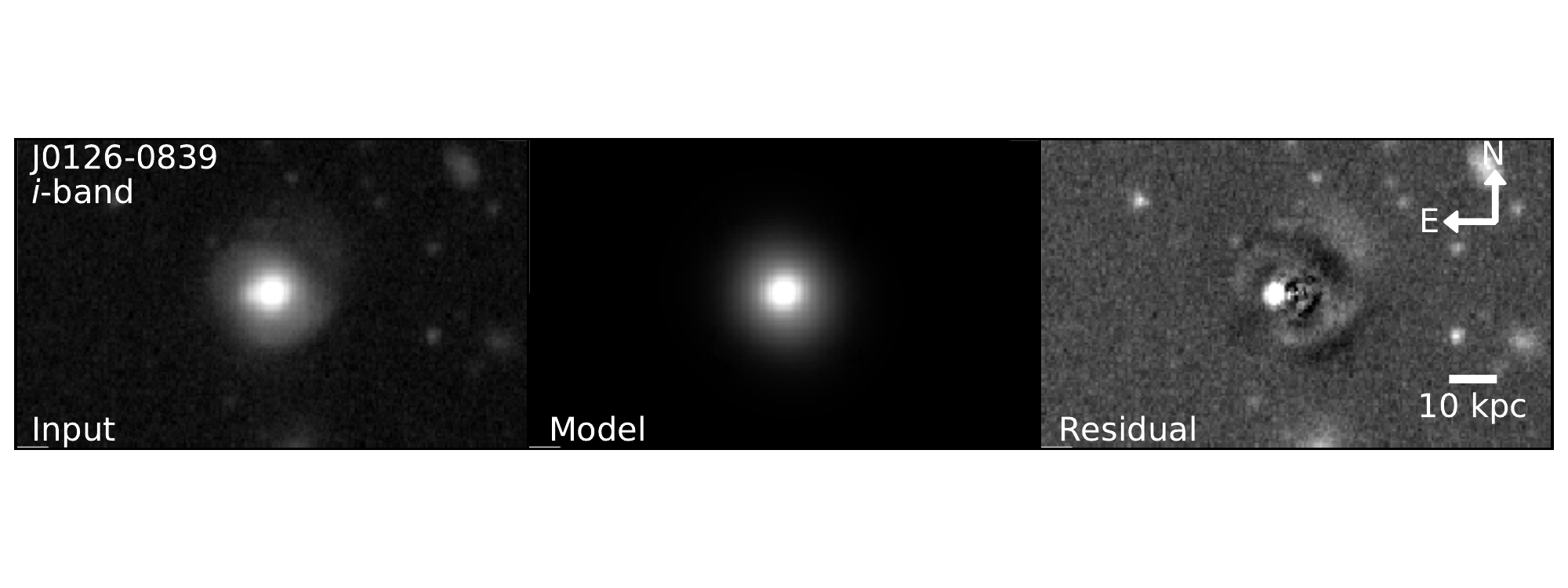}
    \\
    \\
    \includegraphics[width=\textwidth,trim={0 1.8cm 0 1.8cm},clip]{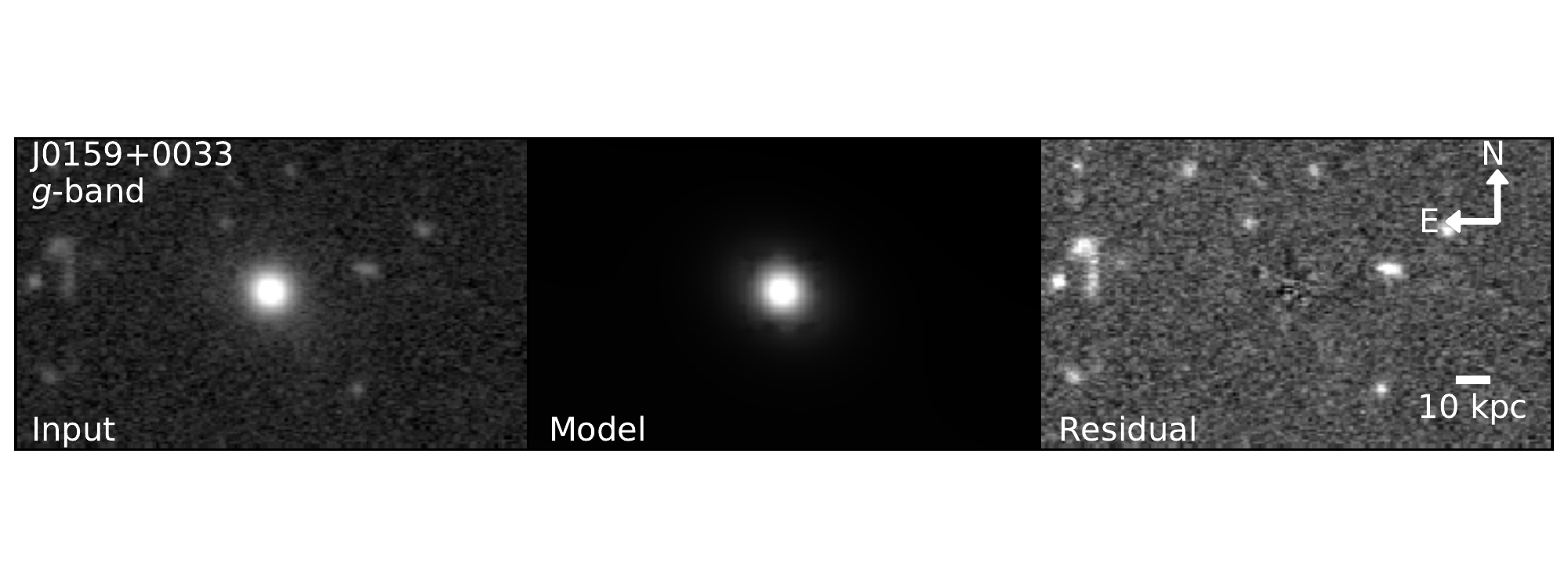}
    \includegraphics[width=\textwidth,trim={0 1.8cm 0 1.8cm},clip]{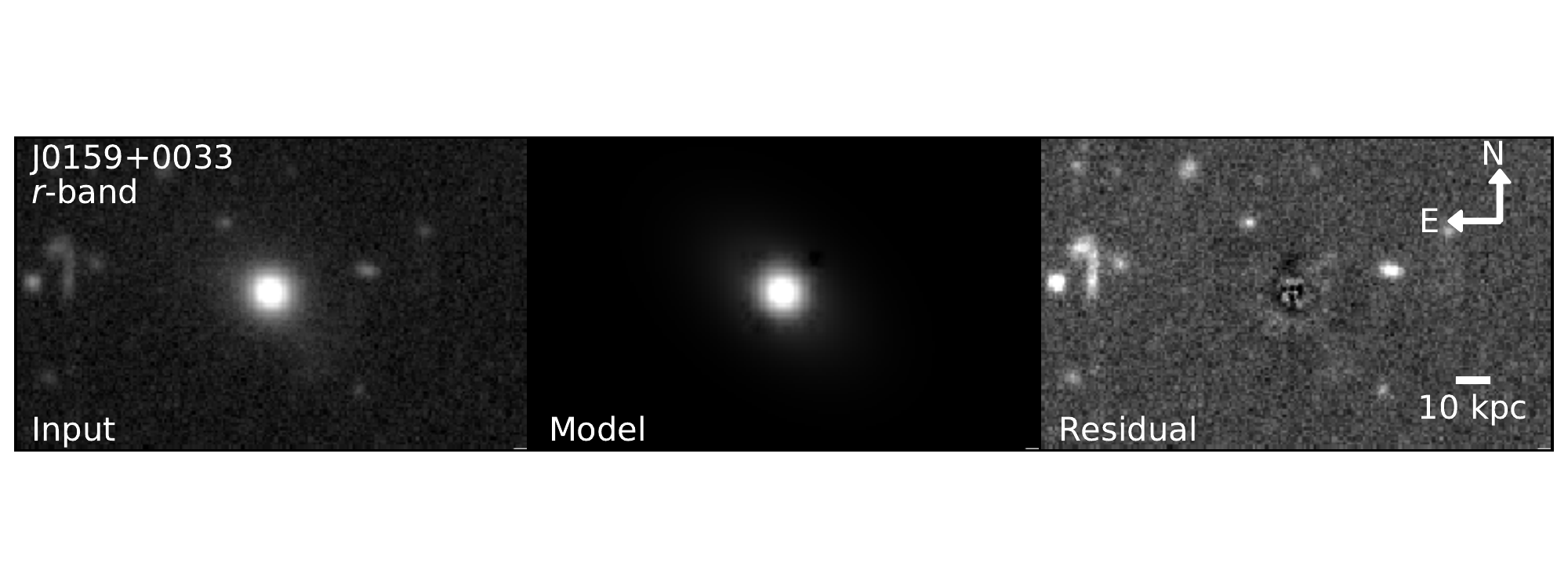}
    \includegraphics[width=\textwidth,trim={0 1.8cm 0 1.8cm},clip]{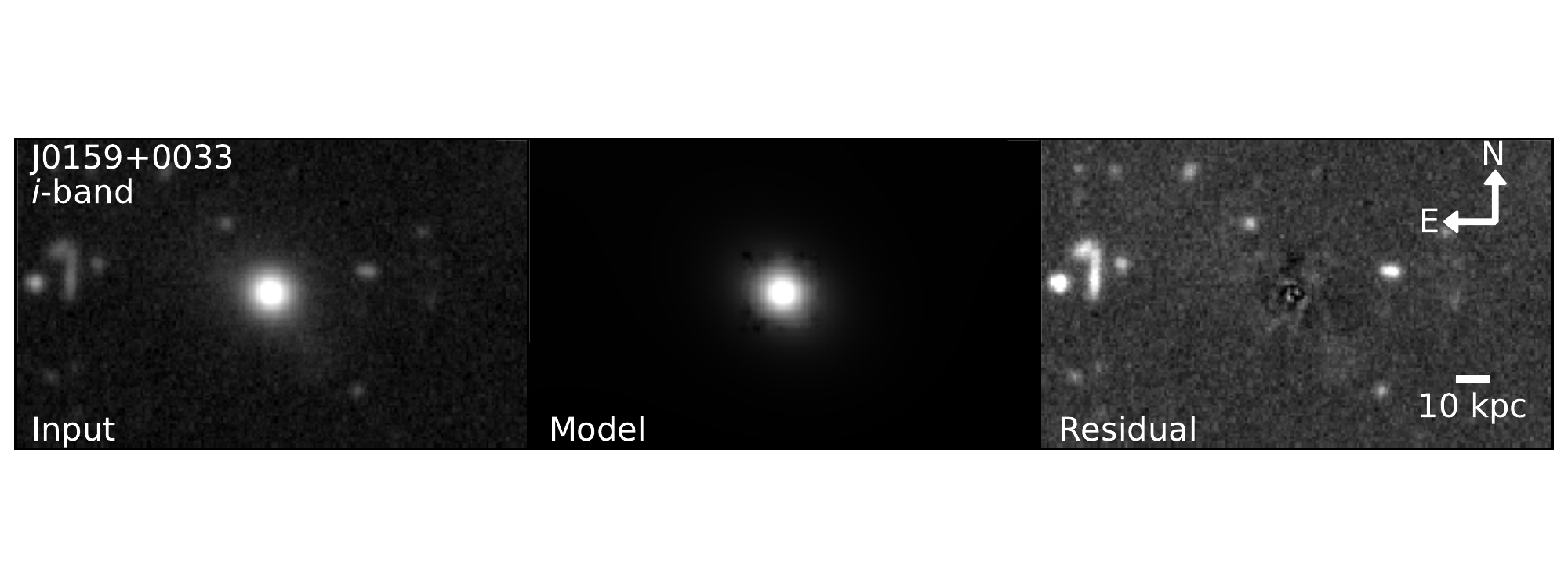}
    \caption{Upper: The data (left), model (middle), and residuals (right) of our J0126$-$0839 fits in \textit{g}, \textit{r}, and \textit{i} (first second and third row, respectively). Lower: As above for J0159+0033}
    \label{fig:J0126_and_J0159}
\end{figure}

\begin{figure}
	\includegraphics[width=\textwidth,trim={0 1.81cm 0 1.8cm},clip]{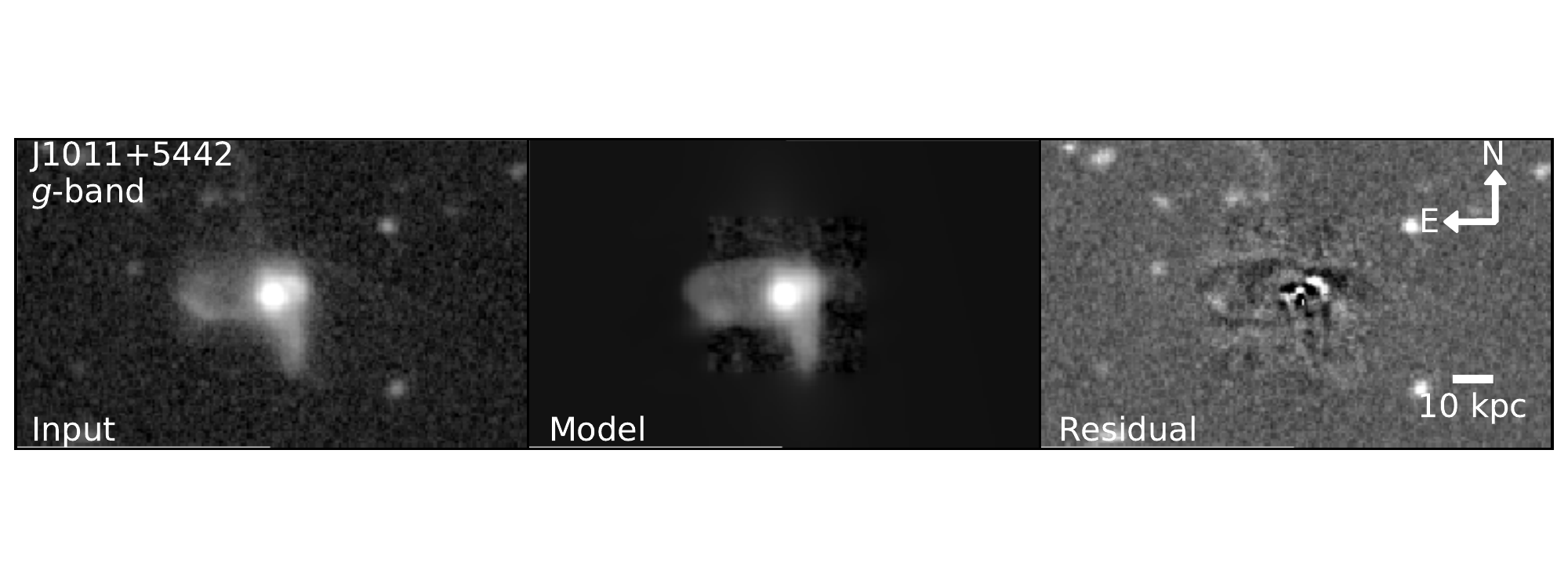}
    \includegraphics[width=\textwidth,trim={0 1.81cm 0 1.8cm},clip]{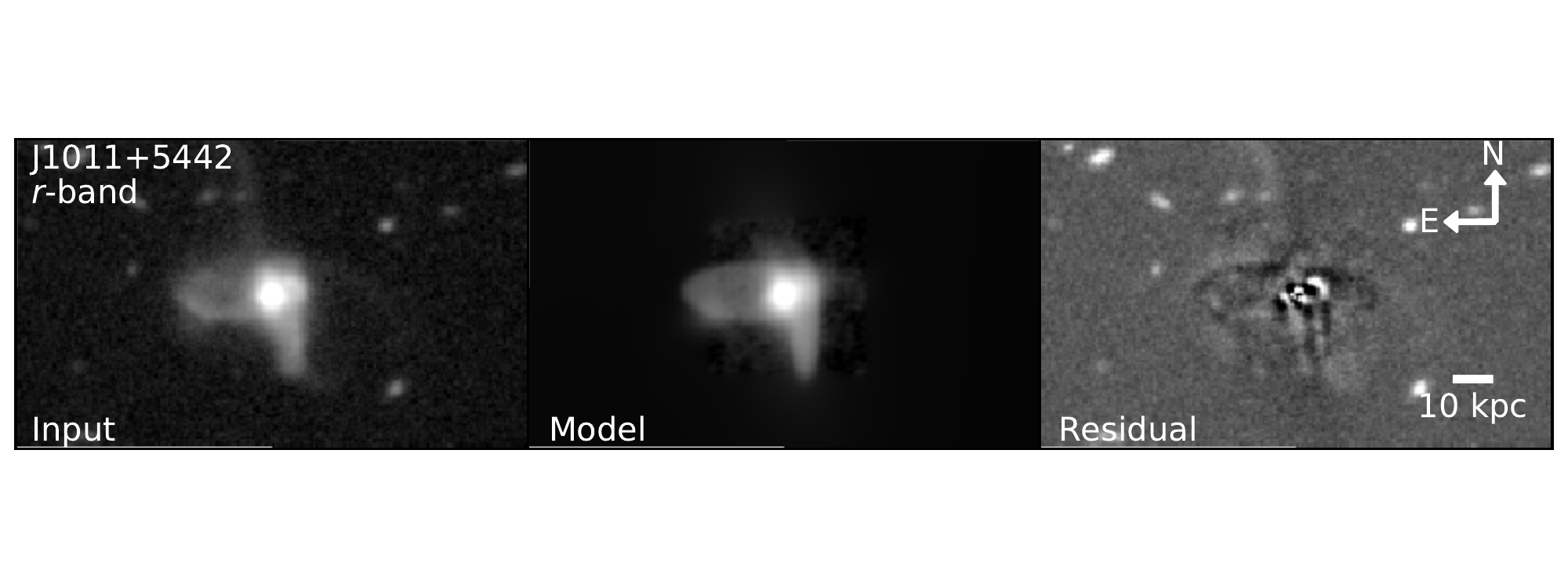}
    \includegraphics[width=\textwidth,trim={0 1.6cm 0 1.8cm},clip]{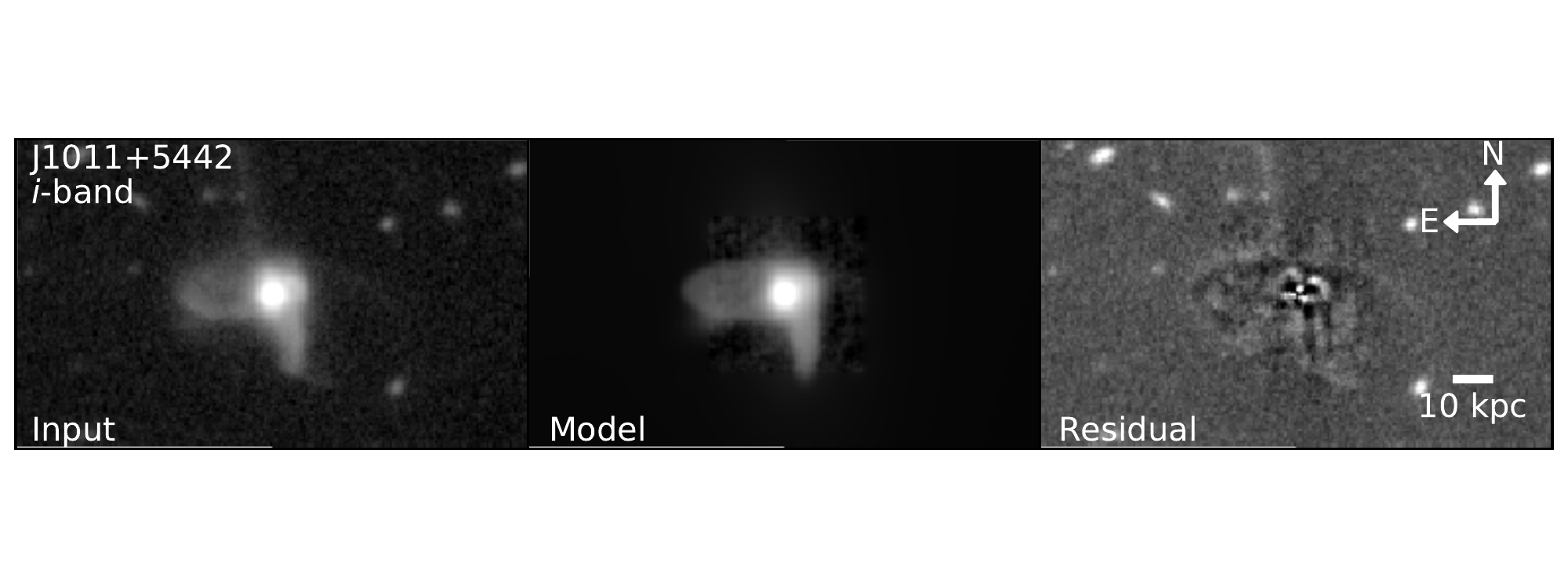}
    \\
    \\
    \includegraphics[width=\textwidth,trim={0 1.8cm 0 1.8cm},clip]{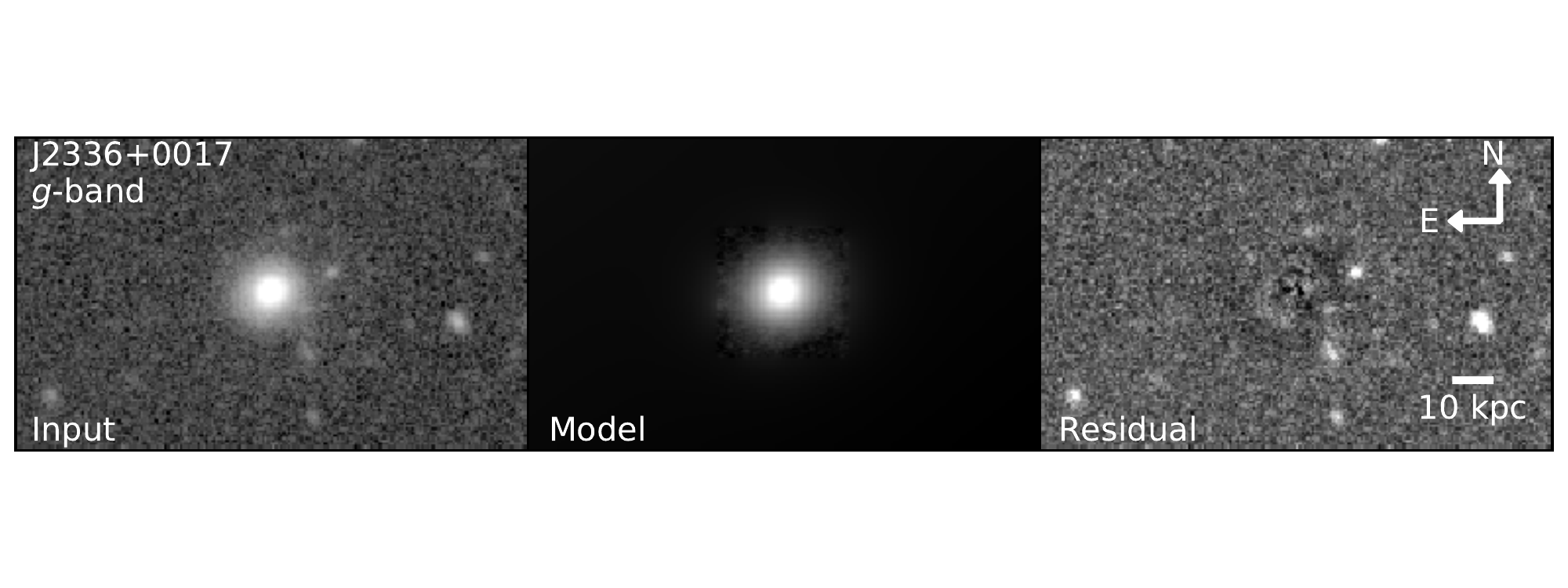}
    \includegraphics[width=\textwidth,trim={0 1.8cm 0 1.8cm},clip]{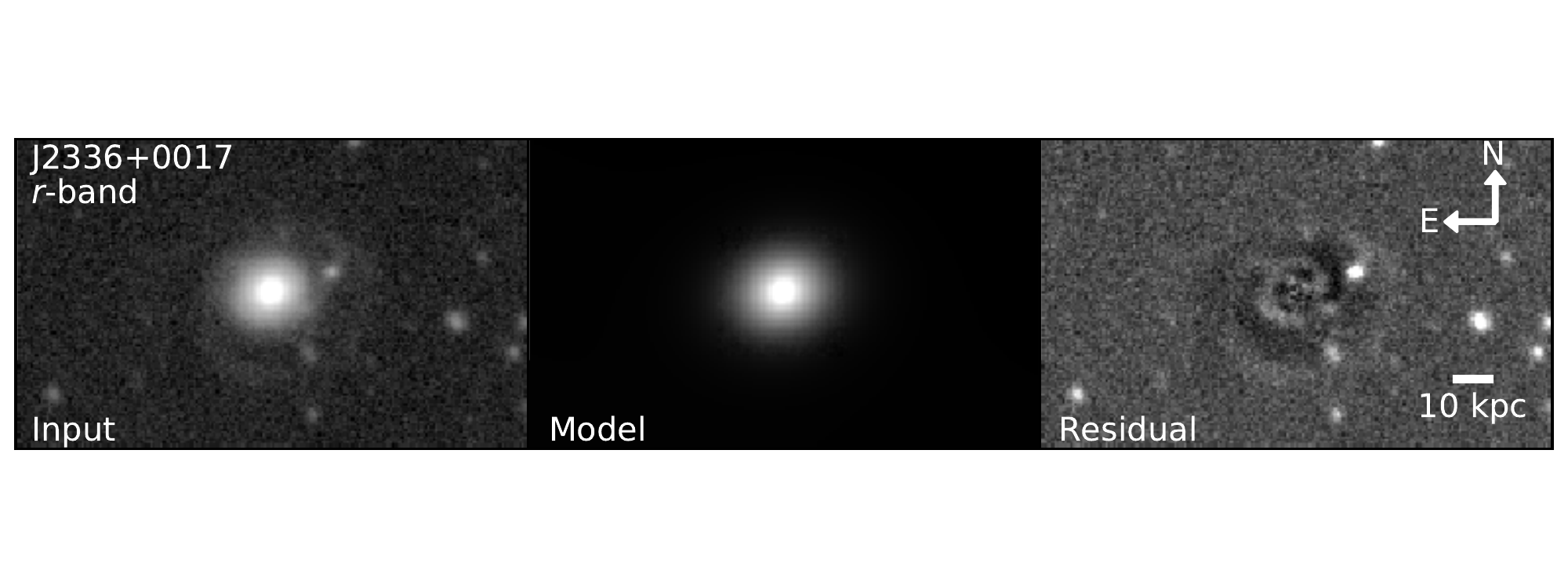}
    \includegraphics[width=\textwidth,trim={0 1.8cm 0 1.8cm},clip]{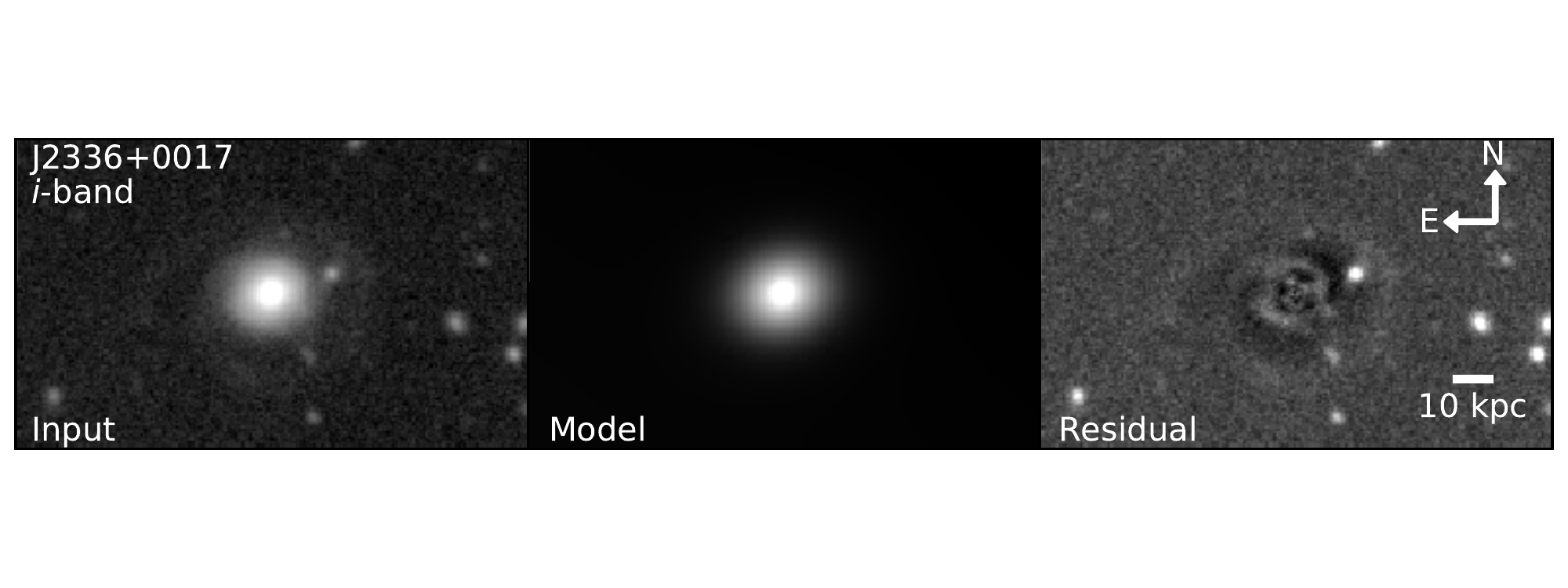}
    \caption{As in Figure \ref{fig:J0126_and_J0159} for J1011+5442 (upper) and J2336+0017 (lower)}
    \label{fig:J1011_and_J2336}
\end{figure}

\end{document}